\documentclass[a4paper,11pt]{article}
\pdfoutput=1 
\usepackage{jheppub} 
\usepackage[T1]{fontenc}
\usepackage{natbib}
\usepackage{soul,xcolor}
\usepackage{diagbox}
\usepackage{mathrsfs,bm}
\usepackage{comment,tcolorbox}
\usepackage{multirow,booktabs,graphicx}
\usepackage{amsmath,amssymb}
\usepackage{hyperref}

\def\d{{\rm d}}
\newcommand{\NN}{\mathcal{N}}
\newcommand{\beq}{\begin{equation}}
\newcommand{\eeq}{\end{equation}}
\newcommand{\bal}{\begin{aligned}}
\newcommand{\eal}{\end{aligned}}

\newtcolorbox{keybox}[1][]{colback=gray!6,colframe=black!60,boxrule=0.5pt,arc=1pt,left=6pt,right=6pt,top=4pt,bottom=4pt,#1}

\title{\boldmath The Love for Three Logarithms: \\
Universal Running of Dynamical Love Numbers}

\author{Sebastian Garcia-Saenz}
\emailAdd{sgarciasaenz@sustech.edu.cn}

\affiliation{
Department of Physics, Southern University of Science and Technology, \\
Shenzhen 518055, China
}

\abstract{
The static tidal Love numbers of black holes vanish in general relativity (GR), but are generically non-zero, and in fact `run' logarithmically with scale, in more general theories or beyond the static limit. We expose a model-agnostic framework for the running of Love numbers at the dynamical level, valid for any tidal perturbation on any static, spherically symmetric, asymptotically flat background, based on the near-zone expansion in powers of the frequency $\omega$ and on the monodromy at spatial infinity. A master recurrence formula determines the monodromy generator to all orders in $\omega^2$, whose three independent entries fully encode the running of the Love numbers. This structure explains the quadratic nature of the renormalization group equation, here proved to all orders and for any type of Love number. We discuss the scheme dependence of the logarithmic coefficients, which is made transparent by the monodromy form. In particular, this clarifies that the anomalous exponents of the tidal field, the so-called `renormalized angular momentum' of the Mano-Suzuki-Takasugi formalism, are scheme-invariant and directly computable on any spacetime, without need of explicit solutions.

We discuss several applications to black holes, including the derivation of a closed-form formula for the leading dynamical running of axial Love numbers of Reissner-Nordstr\"om black holes, proving in particular that it vanishes in the extremal limit. For generic black holes beyond GR, we prove that dynamical running probes deformations of the metric at lower multipoles than the static one, and we illustrate the power of the method by deriving some exact results for the Hayward and Bardeen regular black holes.
}

\begin{document}
\maketitle
\flushbottom

\section{Introduction} \label{sec:intro}

Tidal Love numbers quantify the response of a compact object to an external tidal field and are among the few properties of the object that leave a clean imprint on the gravitational waves emitted by a binary inspiral~\cite{Flanagan:2007ix,Hinderer:2007mb,Damour:2009vw,Binnington:2009bb} (for a review, see~\cite{Rodriguez:2026iot}). For black holes in general relativity (GR) the static Love numbers famously vanish~\cite{Fang:2005qq,Hui:2020xxx,LeTiec:2020spy,LeTiec:2020bos,Chia:2020yla,Kol:2011vg,Charalambous:2021mea,Poisson:2021yau}, a statement which is puzzling from the viewpoint of the world-line effective field theory (EFT)~\cite{Goldberger:2004jt,Goldberger:2007hy,Rothstein:2014sra,Porto:2016pyg,Porto:2016zng}, but can be explained by certain hidden symmetries of the perturbation equations~\cite{Charalambous:2021kcz,Charalambous:2022rre,Hui:2021vcv,Penna:2018gfx,Hui:2022vbh,BenAchour:2022uqo,Combaluzier-Szteinsznaider:2024sgb,DeLuca:2024nih,Berens:2025okm,Parra-Martinez:2025bcu,DeLuca:2025zqr,Sharma:2024hlz,Sharma:2025xii}. This `no-Love' theorem is, however, a rather delicate accident. As soon as one departs from four-dimensional electro-vacuum GR --- higher dimensions~\cite{Kol:2011vg,Hui:2020xxx,Emparan:2017qxd}, EFT-corrected black holes~\cite{Cardoso:2017cfl,Katagiri:2023umb,DeLuca:2022tkm,DeLuca:2023mio,Garcia-Saenz:2025urd,Noumi:2026shc}, non-minimally coupled fields~\cite{Garcia-Saenz:2022wsl} or, as we discuss at length here, non-zero frequency~\cite{Saketh:2023bul,Perry:2023wmm,Ivanov:2024sds,Katagiri:2024fpn,Caron-Huot:2025tlq,Kobayashi:2025vgl,Combaluzier--Szteinsznaider:2025eoc,Chakraborty:2025wvs,Kosmopoulos:2025rfj,Apostolidis:2026qsg,Kehagias:2026opd} --- the Love numbers are generically non-zero and, more importantly, they \emph{run}: the perturbation contains a logarithm of the radial distance, whose coefficient is the main object of this paper.

This `logarithmic Love number' has two faces. On the mathematical side, it is a simple prediction of Fuchsian theory: the growing and decaying solutions of the perturbation equation near spatial infinity have exponents which differ by an integer, so that the Frobenius series of the growing solution generically develops a logarithm at the resonant power. On the EFT side the same logarithm arises from the renormalization of ultraviolet divergences, and its coefficient is nothing but the beta function of the corresponding Wilson coefficient~\cite{Goldberger:2004jt,Kol:2011vg,Ivanov:2022hlo,Ivanov:2022qqt,Mandal:2023hqa,Hadad:2024lsf,Caron-Huot:2025tlq,Kosmopoulos:2025rfj,Barbosa:2025uau,Barbosa:2026qcv,Elkhidir:2026lgp}. The two descriptions agree because the coefficient of the logarithm is, unlike the Love number itself, independent of the boundary condition imposed at the surface of the object: it is fixed by the exterior geometry alone, and it is therefore a \emph{universal} quantity, the same for a black hole, a star or an exotic compact object with the same exterior metric. In the static case this observation was recently exploited to derive a closed-form formula for the coefficient of the logarithm, valid for any metric and any type of perturbation~\cite{Garcia-Saenz:2025urd}. By the same token, this explains why the Love number itself is ambiguous~\cite{Fang:2005qq,Gralla:2017djj,Creci:2021rkz,Katagiri:2024fpn}: the growing solution is defined only up to the addition of a decaying one, and different prescriptions can lead to different numbers depending on the choice of `scheme'.

The dynamical problem makes this aspect all the more interesting, on both sides. At finite frequency black holes have non-zero Love numbers, and logarithmic running, even in GR~\cite{Kobayashi:2025vgl,Combaluzier--Szteinsznaider:2025eoc,Caron-Huot:2025tlq}. On the EFT side, the running of the dynamical tidal couplings was found to obey a quadratic renormalization group equation in the setting of Schwarzschild, with the quadratic term first appearing at $\mathcal{O}(\omega^{2\ell+1})$ for multipole $\ell$~\cite{Caron-Huot:2025tlq,Chang:2026eti,Solon:2026ubm,Correia:2026utp}. On the full-theory side, the frequency-dependent solutions have been obtained for Schwarzschild and Kerr through the Mano-Suzuki-Takasugi (MST) formalism~\cite{Mano:1996vt,Mano:1996mf,Sasaki:2003xr}, in which the `renormalized angular momentum' $\nu_{\rm MST}$ replaces $\ell$ and encodes the true exponents of the solutions near infinity (see also~\cite{Bonelli:2021uvf,Bonelli:2022ten,Bautista:2023sdf,Aminov:2024mul}). These methods are powerful but tailored to specific backgrounds, and the relation between the various logarithms in the tidal and in the response fields and the various schemes has remained somewhat opaque.

The purpose of this paper is to clarify that the entire structure of logarithmic Love numbers is fixed by a single object, the monodromy of the perturbation equation around the singularity at infinity, and that this object can be computed algebraically on any static and spherically symmetric background. Our starting point is the near-zone expansion of the perturbation in powers of $\omega^2$, in which each order is a Fuchsian problem with the same resonant exponents $\ell$ and $-\ell-1$, sourced by the previous order. We write down a master recurrence formula which solves this tower to all orders, logarithms included, in a canonical basis of solutions, referred to here as `unit pure-power scheme'. The central result is that, to all orders in $\omega^2$, the growing and decaying solutions are of the form $x^{\NN}\Phi$, where $\Phi$ is free of logarithms and $\NN(\omega^2)$ is a traceless $2\times2$ matrix, the monodromy generator, whose three independent entries $A$, $B$ and $C$ are power series in $\omega^2$. We refer to these as the \emph{three logarithms}: $A$ measures the running of the Love number, $C$ is identified with the anomalous dimension of the tidal field, and $B$ is interpreted as a tail-like term. All the logarithms of the near-zone solution, at every order and to every power, are generated by the function $r^\NN$, with some immediate consequences:
\begin{itemize}
	\item The logarithms are resummed, and the exact exponents of the solutions near infinity are those of $u_g\sim r^{\nu_{\rm MST}+1}$ and $u_d\sim r^{-\nu_{\rm MST}}$ with $\nu_{\rm MST}=\ell-\delta$ and $\delta^2=C^2-AB$, where the former is nothing but the MST renormalized angular momentum, here shown to be computable by a linear algebraic recursion, on any background and with no need of solving any differential equation;
	\item A change of the reference scale in the logarithms operates as a M\"obius transformation on the Love number, resulting in the renormalization group equation $\tfrac{\partial\mathcal{K}}{\partial\log\mu}=-A-2C\mathcal{K}-B\mathcal{K}^2$, valid to all orders in the expansion in $\omega$ and for any spacetime and any tidal field;
	\item The monodromy generator transforms covariantly under changes of basis, which makes the scheme dependence of the logarithmic Love numbers transparent; this in particular establishes a dictionary between different renormalization schemes.
\end{itemize}

The second half of the paper is devoted to applications to black holes. For Schwarzschild we recover the known results for the leading dynamical running of scalar, vector and gravitational perturbations, derived here in a single closed-form formula. As additional checks, we compute $\nu_{\rm MST}$ through $\mathcal O(\omega^4)$ and the renormalization group equation through $\mathcal O(\omega^6)$, both in agreement with the results of the MST and EFT literatures, respectively. For Reissner-Nordstr\"om (RN), for which dynamical Love numbers have not been studied (except in five dimensions~\cite{Cvetic:2026wht}), we obtain the leading dynamical running of axial electromagnetic and gravitational perturbations in closed form for all multipoles. Interestingly, we discover that it vanishes identically in the extremal limit. We also derive several general results valid for arbitrary metrics. For instance, we prove that the leading logarithmic tail coefficient is universal, the same for any black hole and tidal field. We further illustrate the advantages of the method by presenting exact results for the Hayward and Bardeen regular black holes.

It is worth mentioning that the study of monodromy in the context of black hole perturbations is not new~\cite{Castro:2013kea,Castro:2013lba,DiRusso:2024hmd,Aminov:2024mul,Aggarwal:2019iay,Chanson:2020hly,Pradhan:2026czt}, although previous works were mainly concerned with the monodromies at the black hole horizons rather than at the irregular singularity at spatial infinity. Still, some of the results we derive here are known from studies of specific black hole backgrounds. In particular~\cite{Castro:2013lba} identified and computed the MST renormalized angular momentum for a scalar field on Schwarzschild and Kerr, while~\cite{Aminov:2024mul} obtained the monodromy generator around infinity for the Regge-Wheeler equation and understood the resulting M\"obius structure of its elements. What our method adds is that the full generator, and not only its eigenvalue, is computable for any background and any tidal field via an explicit algebraic recurrence that requires no knowledge of the static solutions; that its entries are in direct correspondence with the logarithmic Love numbers, with a direct physical meaning; and that its covariance under changes of basis precisely organizes the scheme dependence.

The paper is organized as follows. Sec.\ \ref{sec:monodromy} starts with our set-up of the near-zone expansion and establishes a master recurrence formula that determines the structure of the growing and decaying solutions. We then prove our main `three logarithms' theorem, which we next relate with the renormalized angular momentum and the question of scheme (in)dependence. Sec.\ \ref{sec:renormalization} derives the M\"obius transformation relating the Love number at different scales, which we use to prove the exact form of the renormalization group equation. Sec.\ \ref{sec:black holes} contains applications to black holes, including: general results for Schr\"odinger-type equations, the Regge-Wheeler equation, the Moncrief equations and the Klein-Gordon equation on general backgrounds. We conclude in Sec.\ \ref{sec:conclusion} with some final remarks. Three appendices contain some technical derivations and proofs.

\bigskip

\textbf{Set-up and notations:} We focus exclusively on four-dimensional, asymptotically flat, static and spherically symmetric backgrounds. For the radial coordinate we use both the areal radius $r$ and the variable $x:= r_g/r$, where $r_g:=2GM\equiv 1$ is the Schwarzschild radius defined in terms of the ADM mass $M$, so that the metric functions $-g_{tt},g^{rr}=1-x+\cdots$ at spatial infinity.

The equations for perturbations are assumed to have been already separated in terms of spherical harmonics. Throughout we omit the harmonic number labels $\ell,m$ on all functions and coefficients.


\section{Near-zone monodromy} \label{sec:monodromy}

\subsection{Near-zone expansion}

Throughout this paper we will be studying ODEs of the form
\beq \label{eq:general ode}
\hat{L}_0u+\omega^2 \mathcal{R}(x)u=0 \,,\qquad \hat{L}_0:=x^2\frac{\d^2}{\d x^2}+xp(x)\frac{\d}{\d x}+q(x) \,,
\eeq
where $p$ and $q$ are assumed analytic at $x=0$, while $\mathcal{R}(x)$ admits a Laurent expansion starting with the power $x^{-2}$,
\beq\bal
p=\sum_{n\geq0}p_nx^n \,,\qquad q=\sum_{n\geq0}q_nx^n \,,\qquad \mathcal{R}=x^{-2}\sum_{n\geq0}w_nx^n \,.
\eal\eeq
It follows that $x=0$ is an irregular singular point of \eqref{eq:general ode}. All other singularities of the equation are assumed to lie at $|x|\gtrsim1$ on the complex plane. The exception is when the parameter $\omega=0$, in which case $x=0$ is a regular singularity. Solutions are then characterized by the roots of the indicial polynomial $I(s)$, i.e.\ the Frobenius exponents $s_1$ and $s_2$. For four-dimensional static and spherically symmetric systems,
\beq
I(s):=(s-\ell)(s+\ell+1) \,,\qquad \ell=0,1,2,\ldots \,,
\eeq
so that $s_1-s_2=2\ell+1$, i.e.\ the case of Frobenius resonance, implying (generically) the presence of a logarithm in the solution. This is the well-known running term of static Love numbers.

We suppose $\omega\ll1$, so that in the \textit{near-zone} region, $\omega\ll x\lesssim 1$, we can treat the irregular term in \eqref{eq:general ode} as a perturbation, leading to the (possibly asymptotic) expansion
\beq
u(x;\omega)=\sum_{k\geq0}\omega^{2k}u^{(2k)}(x) \,,\qquad \hat{L}_0u^{(0)}=0 \,,\qquad \hat{L}_0u^{(2k)}=-\mathcal{R}u^{(2k-2)} \,.
\eeq
At each \textit{level} $k$, the task is therefore to solve a \textit{sourced} Fuchsian problem with the same resonant exponents $\ell$ and $-\ell-1$. From the general theory of this type of problems (see e.g.~\cite{wasow1987asymptotic}), we know that the solutions can be expressed in powers series times, for each power, a polynomial in $\log x$. At level $k$, the leading powers are $x^{\ell-2k}$ and $x^{-\ell-1-2k}$ (since $\mathcal{R}\sim x^{-2}$, at each level the source's singularity strength increases by 2), corresponding respectively to the decaying and growing solutions,
\beq\bal
u_{\rm d}&=\sum_{k\geq0}\omega^{2k}x^{\ell-2k}\sum_{n=0}^{\infty}\sum_{j=0}^{2k+1}D_{k,n,j}x^n\log^jx \,,\\
u_{\rm g}&=\sum_{k\geq0}\omega^{2k}x^{-\ell-1-2k}\sum_{n=0}^{\infty}\sum_{j=0}^{2k+1}G_{k,n,j}x^n\log^jx \,.
\eal\eeq

At each level, resonances occur whenever $n$ is such that the power matches the Frobenius exponents: $n=2k$ and $n=2k\pm(2\ell+1)$. Substituting into the level-$k$ equation produces the following master recurrence formula:
\begin{keybox}[title={Master recurrence formula}]
	Let $E\in\{D,G\}$, $s_0\in\{-\ell-1,\ell\}$ and $s=s_0-2k+n$. For $k=0,1,2,\dots$, for $n=0,1,2,\dots$, and for $j=2k+1,2k,\dots,1,0$, \emph{in this order}, $E_{k,n,j}$ is determined by
	\begin{equation}
		\begin{aligned}
			&I(s)E_{k,n,j}+(j+1)I'(s)E_{k,n,j+1}+(j+1)(j+2)E_{k,n,j+2}\\
			&+\sum_{m=1}^{n}\Big\{\big[(s-m)p_m+q_m\big]E_{k,n-m,j}+(j+1)p_mE_{k,n-m,j+1}\Big\}
			+\sum_{m=0}^{n}w_mE_{k-1,n-m,j}=0 \,,
		\end{aligned}
		\label{eq:master-recursion}
	\end{equation}
	and $E_{k,n,j}=0$ whenever $k<0$, $n<0$, $j<0$ or $j>2k+1$.
\end{keybox}
To apply this formula, one distinguishes between resonant and non-resonant powers:
\begin{itemize}
	\item \textit{Non-resonant case:} if $I(s)\neq0$ the master formula determines $E_{k,n,j}$.
	\item \textit{Resonant case:} if $I(s)=0$ the master formula determines $E_{k,n,j+1}$. Computing through $j=2k,2k-1,\ldots,0$ leaves $E_{k,n,0}$ undetermined, but this multiplies a homogeneous solution and may be chosen as zero (except for $k=0=n$).\footnote{To see that $j>2k+1$ cannot appear, observe that the source at level $k$ has the log-powers of level $k-1$. At each level, the log-power can increase by at most 2, since there are only two resonances. The maximum log-power thus satisfies $J_k=J_{k-1}+2$, and we know that $J_0=1$. Therefore $J_k=2k+1$. Notice that $J_k$ is only an upper bound; we will see for instance that at level $k=1$ cubic powers do not appear.\label{fn:log level}}
\end{itemize}
We thus adopt a `pure-power' normalization convention in which the growing solution contains $x^{-\ell-1}$ with coefficient $1$ and $x^{\ell}$ with coefficient $0$ (apart from logarithms) at every order in $\omega^2$, and conversely for the decaying solution. Explicitly, $G_{k,2k,0}=0$ if $k\ge1$, $G_{k,2\ell+1+2k,0}=0$ if $k\ge0$, $D_{k,2k,0}=0$ if $k\ge1$ and $D_{k,2k-2\ell-1,0}=0$ if $k\ge\ell+1$. At level $0$ we normalize $G_{0,0,0}=1=D_{0,0,0}$.

The \textit{physical} solution, i.e.\ the linear combination consistent with the boundary condition at the black hole horizon, is given by
\beq \label{eq:uphys}
u_{\rm phys}=u_{\rm g}+\mathcal{K}(\omega)u_{\rm d} \,,
\eeq
for some constant $\mathcal{K}(\omega)$ that we identify as the Love number. Fuchsian theory makes it obvious that the non-analyticities at $x=0$, i.e.\ the logarithms, are `universal', in the sense that they are independent of boundary conditions. They are (generically) present in $u_{\rm phys}$ whatever the value of $\mathcal{K}(\omega)$, and are therefore the same for every object with the same exterior metric. From this perspective, `logarithmic Love number' is a somewhat misleading name: it is not a Love number, but rather the running of the Love number; see Sec.\ \ref{sec:renormalization}.

It is instructive to work out the first two levels explicitly before we discuss some general properties.

\textbf{Level $k=0$ (static).} We write $G_{0,n,0}\equiv b_n$ and $G_{0,n,1}\equiv \tilde b_n$, with $b_0=1$ and $\tilde b_0=0$ as per our normalization. For each $n$, the master formula \eqref{eq:master-recursion} has two log-power levels:
\begin{align}
	j=1:&\quad n(n-2\ell-1)\tilde b_n+\sum_{m=1}^{n}\big[(n-\ell-1-m)p_m+q_m\big]\tilde b_{n-m}=0 \,,\\
	j=0:&\quad n(n-2\ell-1)b_n+(2n-2\ell-1)\tilde b_n \nonumber\\
	&\quad+\sum_{m=1}^{n}\Big\{\big[(n-\ell-1-m)p_m+q_m\big]b_{n-m}+p_m\tilde b_{n-m}\Big\}=0 \,.
\end{align}
From the $j=1$ level we infer that $\tilde b_n=0$ for $1\le n\le2\ell$. The $j=0$ level then yields
\begin{equation}
	b_n=-\frac{1}{n(n-2\ell-1)}\sum_{m=1}^{n}\big[(n-\ell-1-m)p_m+q_m\big]b_{n-m} \,,\qquad 1\le n\le2\ell \,.
\end{equation}
At $n=2\ell+1$ we encounter a resonance ($s=\ell$); the $j=1$ line is empty and the $j=0$ line determines the static logarithmic Love number (with $b_{2\ell+1}=0$ by our pure-power normalization), i.e.\ the coefficient of $-x^{\ell}\log x$ in the solution:\footnote{The minus sign is so that the logarithmic Love number matches the coefficient of $r^{-\ell}\log r$, as is usually done in the literature.}
\begin{equation}\bal
	\alpha_0:= -\tilde b_{2\ell+1}&=\frac{1}{2\ell+1}\sum_{m=1}^{2\ell+1}\big[(\ell-m)p_m+q_m\big]b_{2\ell+1-m}\\
	&=\frac{1}{2\ell+1}\sum_{m=0}^{2\ell}\big[(m-\ell-1)p_{2\ell+1-m}+q_{2\ell+1-m}\big]b_{m} \,.
	\label{eq:a0master}
\eal\end{equation}
The second line is just a rewriting of the first, and is precisely the form derived in~\cite{Garcia-Saenz:2022wsl}. For $n>2\ell+1$ the above formulae continue to determine $\tilde b_n$ and $b_n$ recursively. On the other hand, the decaying branch does not exhibit any resonances and no logarithm appears in $u_{\rm d}$ at the static level. Letting $\tilde d_n\equiv D_{0,n,0}$, we have $\tilde d_n=-\frac{1}{n(n+2\ell+1)}\sum_{m=1}^{n}[(\ell+n-m)p_m+q_m]\tilde d_{n-m}$, all determined recursively in terms of $\tilde d_0\equiv 1$.

In fact, the $\tilde b_n$ and $\tilde d_n$ coefficients are related. We separate $u_{\rm g}\equiv \Phi+\Psi\log x$. A direct calculation shows that $\hat{L}_0\Psi=0$,\footnote{Proof: Since $\hat{L}_0$ is linear we have $0=\hat{L}_0u_{\rm g}=\hat{L}_0\Phi+\hat{L}_0(\Psi\log x)=\hat{L}_0\Phi+(\hat{L}_0\Psi)\log x+2x\Psi'+(p-1)\Psi$. The analytic and non-analytic parts of the last expression must separately vanish, i.e.\ $\hat{L}_0\Psi=0$.} i.e.\ $\Psi$ is a static solution, which by definition does not contain logarithms. It must therefore be proportional to $u_{\rm d}$. The coefficient of $x^{\ell}$ in $\Psi$ is $\tilde b_{2\ell+1}\equiv -\alpha_0$, while the coefficient of $x^{\ell}$ in $u_{\rm d}$ is $1$. Therefore $\Psi=-\alpha_0u_{\rm d}$ or, matching powers, $\tilde b_{2\ell+1+n}=-\alpha_0\tilde e_n$. This result is a particular case of a general theorem that we discuss below.

\textbf{Level $k=1$ (LO dynamical).} We write $G_{1,n,0}\equiv h_n$, $G_{1,n,1}\equiv \tilde{h}_n$ and $G_{1,n,2}\equiv \hat{h}_n$. For each $n$, the master formula \eqref{eq:master-recursion} has three log-power levels:\footnote{The $j=3$ line, $I(s)C_{1,n,3}+\sum_{m=1}^{n}\big[(s-m)p_m+q_m\big]C_{1,n-m,3}=0$, gives $C_{1,n,3}=0$ for all non-resonant $n$, while the $j=2$ line yields the same result for the resonant $n$; cf.\ also footnote \ref{fn:log level}.}
\begin{align}
	j=2:&\quad I(s)\hat{h}_n+\sum_{m=1}^{n}\big[(s-m)p_m+q_m\big]\hat{h}_{n-m}=0 \,,\label{eq:L1j2}\\
	j=1:&\quad I(s)\tilde{h}_n+2I'(s)\hat{h}_n \nonumber\\
	&\quad+\sum_{m=1}^{n}\Big\{\big[(s-m)p_m+q_m\big]\tilde{h}_{n-m}+2p_m\hat{h}_{n-m}\Big\}+\sum_{m=0}^{n}w_m\tilde b_{n-m}=0 \,,\label{eq:L1j1}\\
	j=0:&\quad I(s)h_n+I'(s)\tilde{h}_n+2\hat{h}_n \nonumber\\
	&\quad+\sum_{m=1}^{n}\Big\{\big[(s-m)p_m+q_m\big]h_{n-m}+p_m\tilde{h}_{n-m}\Big\}+\sum_{m=0}^{n}w_m b_{n-m}=0 \,.\label{eq:L1j0}
\end{align}
We start with $n=0,1$, which are non-resonant. From \eqref{eq:L1j2} we infer $\hat{h}_{0,1}=0$, and then from \eqref{eq:L1j1} we infer $\tilde{h}_{0,1}=0$. At the first resonance, $n=2$, \eqref{eq:L1j1} yields $\hat{h}_2=0$ and \eqref{eq:L1j0} yields
\begin{equation} \label{eq:delta1 recursion}
	\boxed{\ \gamma_1:= \tilde{h}_2=\frac{1}{2\ell+1}\left\{\sum_{m=1}^{2}\big[(-\ell-1-m)p_m+q_m\big]h_{2-m}+\sum_{m=0}^{2}w_mb_{2-m}\right\}\ }
\end{equation}
Notice that this formula depends only on $h_0$ and $h_1$ ($h_2=0$ by our normalization) and the level-0 terms $b_{0,1,2}$, a fact which will allow us to derive a closed-form expression when we specify to black hole metrics in Sec.\ \ref{sec:black holes}.

At the second resonance, $n=2\ell+3$, we find $\hat{h}_{2\ell+3}=0$ from \eqref{eq:L1j1} (which may also be inferred directly from the theorem below), while from \eqref{eq:L1j0} we get
\begin{equation} \label{eq:alpha1 recursion}
	\boxed{\ \alpha_1:=-\tilde{h}_{2\ell+3}=\frac{1}{2\ell+1}\left\{\sum_{m=1}^{2\ell+3}\Big[\big[(\ell-m)p_m+q_m\big]h_{2\ell+3-m}+p_m\tilde{h}_{2\ell+3-m}\Big]+\sum_{m=0}^{2\ell+3}w_mb_{2\ell+3-m}\right\}\ }
\end{equation}

Formulae \eqref{eq:delta1 recursion} and \eqref{eq:alpha1 recursion} are the $\mathcal{O}(\omega^2)$ analogues of \eqref{eq:a0master}, i.e.\ the coefficients of $x^{-\ell-1}\log x$ and $-x^{\ell}\log x$ in the solution $u_{\rm g}$, respectively (with the former equal to zero at level 0). For the decaying branch, there is a resonance at $n=2$, which yields the coefficient of $x^{\ell}\log x$ in $u_{\rm d}$: $\bar{\gamma}_1\equiv D_{1,2,1}=-\gamma_1$, which is again a consequence of the theorem below. In the particular case $\ell=0$, a second resonance occurs at $n=1$, and one finds the result
\beq \label{eq:beta1 result}
\beta_1:= D_{1,1,1}^{(\ell=0)}=2 \,,
\eeq
valid for \textit{any} black hole metric and (scalar) tidal perturbation field.\footnote{Since this is only valid for $\ell=0$, it is not applicable to vector and tensor perturbations. Notice however that here we are not assuming the Klein-Gordon equation (although we will specify to this case in Sec.\ \ref{sec:black holes}), so \eqref{eq:beta1 result} would remain valid for a non-minimally coupled scalar, for instance.} This is in fact a particular case of a general result we derive in Sec.\ \ref{sec:black holes}.

\subsection{Three logarithms}

We define the following four series of resonant logarithmic coefficients:
\begin{equation}
	\begin{aligned}
		A(\omega^2)&:=-\sum_{k\ge0}G_{k,\,2\ell+1+2k,\,1}\,\omega^{2k}=\alpha_0+\alpha_1\omega^2+\alpha_2\omega^4+\cdots \,,\\
		C(\omega^2)&:=\sum_{k\ge1}G_{k,\,2k,\,1}\,\omega^{2k}=\gamma_1\omega^2+\gamma_2\omega^4+\cdots \,,\\
		B(\omega^2)&:=\sum_{k\ge\ell+1}D_{k,\,2k-2\ell-1,\,1}\,\omega^{2k}=\beta_{\ell+1}\omega^{2\ell+2}+\cdots \,,\\
		\bar C(\omega^2)&:=\sum_{k\ge1}D_{k,\,2k,\,1}\omega^{2k} \,.
	\end{aligned}
	\label{eq:ADBdef}
\end{equation}
These are, respectively, the coefficients of $-x^{\ell}\log x$ in $u_{\rm g}$, $x^{-\ell-1}\log x$ in $u_{\rm g}$, $x^{-\ell-1}\log x$ in $u_{\rm d}$ and $x^{\ell}\log x$ in $u_{\rm d}$. Obviously, the near-zone solutions contain many more logarithmic terms than these four series (all the $\tilde h_n$ coefficients at level 1, for instance, as well as higher powers of $\log x$). However, the following theorem states that \textit{all} these coefficients are determined by $A$, $B$ and $C$.

\textbf{Theorem.} Let $\Phi_{\rm g}$ and $\Phi_{\rm d}$ be the pure-power parts ($j=0$) of $u_{\rm g}$ and $u_{\rm d}$, respectively. Then, as formal power series in $\omega^2$,
\begin{equation}
	\begin{pmatrix}u_{\rm g}\\ u_{\rm d}\end{pmatrix}=x^{\NN}\begin{pmatrix}\Phi_{\rm g}\\ \Phi_{\rm d} \end{pmatrix} \,,\qquad
	x^{\NN}\equiv e^{\NN\log x} \,,\qquad
	\NN(\omega^2):=\begin{pmatrix}C&-A\\B&-C\end{pmatrix},\qquad \bar C=-C \,.
	\label{eq:theorem}
\end{equation}
Equivalently, the coefficient of $\frac{1}{j!}\log^jx$ in $\binom{u_{\rm g}}{u_{\rm d}}$ is $\NN^j\binom{\Phi_{\rm g}}{\Phi_{\rm d}}$, and analytic continuation once around $x=0$ acts on the basis as $\binom{u_{\rm g}}{u_{\rm d}}\mapsto e^{2\pi i\NN}\binom{u_{\rm g}}{u_{\rm d}}$. In particular $\NN^2=\delta^2\mathbf 1$ with $\delta^2:=C^2-AB$, so that $x^{\NN}=\cosh(\delta\log x)\mathbf 1+\delta^{-1}\sinh(\delta\log x)\NN$.

For the proof, we proceed in steps:\footnote{This theorem is essentially textbook material; see e.g.\ \cite{wasow1987asymptotic}. The only novelty here is perhaps its application to a tower of inhomogeneous equations.}
\begin{itemize}
	\item[]\textit{(i) Monodromy:} At each level $k$, $u_{\rm g,d}$ is a finite sum of terms of the form $x^{\sigma}\log^jx$ times a series which converges for $|x|\lesssim1$ (recall that we assume the nearest singularity of the equation lies at $|x|\gtrsim1$). This is nothing but Fuchs' theorem at the homogeneous level 0, while the conclusion remains at higher levels because the sources, by construction of the perturbative expansion in $\omega^2$, share the same property.
	
	Consider the analytic continuation once around $x=0$ along a loop $\gamma$, which in the present case amounts to the operation $\gamma:\; \log x\mapsto\log x+2\pi i$. Let $U:=\binom{u_{\rm g}}{u_{\rm d}}=\sum_{k\geq0}\omega^{2k}U^{(k)}$. We claim that there exist constant matrices $M_0,M_1,M_2,\ldots$ such that
	\beq \label{eq:monodromy proof1}
	\gamma U^{(k)}=\sum_{m=0}^kM_mU^{(k-m)} \,,
	\eeq
	for all $k$. We prove this by induction. The claim holds at level $k=0$ because the equation's coefficients are Laurent series, thus $\hat{L}_0(\gamma U^{(0)})=\gamma(\hat{L}_0U^{(0)})=0$, i.e.\ $\gamma U^{(0)}=M_0U^{(0)}$ since $U^{(0)}$ is a basis of solutions. We now assume it holds for all levels below $k$, and thus
	\beq
	\hat{L}_0(\gamma U^{(k)})=\gamma(\hat{L}_0U^{(k)})=-\gamma(\mathcal{R} U^{(k-1)})=-\mathcal{R}\gamma U^{(k-1)}=-\mathcal{R}\sum_{m=0}^{k-1}M_m U^{(k-1-m)} \,.
	\eeq
	On the other hand, the candidate combination $V^{(k)}:=\sum_{m=0}^{k-1}M_mU^{(k-m)}$ satisfies
	\beq
	\hat{L}_0V^{(k)}=-\mathcal{R}\sum_{m=0}^{k-1}M_m U^{(k-1-m)} \,,
	\eeq
	the same right-hand side. Hence $\hat{L}_0(\gamma U^{(k)}-V^{(k)})=0$, so $\gamma U^{(k)}-V^{(k)}=M_kU^{(0)}$ for some constant matrix $M_k$. This proves \eqref{eq:monodromy proof1} at level $k$.
	
	It follows that $\gamma U=M(\omega^2)U$ with $M:=\sum_{k\geq0}\omega^{2k}M_k$. Indeed, by the Cauchy product formula,
	\beq
	MU=\sum_{m,m'\geq0}\omega^{2(m+m')}M_mU^{(m')}=\sum_{k\geq0}\omega^{2k}\sum_{l=0}^kM_lU^{(k-l)} =\gamma U \,.
	\eeq
	
	
	The matrix $\NN:=\frac{1}{2\pi i}\log M$ is nothing but the monodromy generator. In particular, at level 0 or $\omega=0$, we have $u_{\rm g}\mapsto u_{\rm g}-2\pi i\alpha_0u_{\rm d}$, $u_{\rm d}\mapsto u_{\rm d}$, so $M(0)=\mathbf 1+2\pi i\NN_0$ with $\NN_0=\big(\begin{smallmatrix}0&-\alpha_0\\0&0\end{smallmatrix}\big)$, $\NN_0^2=0$. We remark that $\NN$ is a well-defined quantity. Indeed, let $M\equiv \mathbf{1}+X$, and we have $X=2\pi i\NN_0+\mathcal{O}(\omega^2)$. Since $\NN_0^2=0$, any product of $j$ factors of $X$ contains at least $\lfloor j/2\rfloor$ factors of order $\omega^2$, i.e.\ $X^j=\mathcal{O}(\omega^{2\lfloor j/2\rfloor})$. Thus, at any given order in $\omega^2$, the series $\log(\mathbf 1+X)=X-\tfrac12X^2+\cdots$ terminates and is therefore well defined. At all orders in $\omega^2$ it is defined at least as a formal power series, which is all we claim.
	
	\item[]\textit{(ii) Normal form:} We observe that $\Phi:=\binom{\Phi_{\rm g}}{\Phi_{\rm d}}= x^{-\NN}\binom{u_{\rm g}}{u_{\rm d}}$ is single-valued: analytical continuation along a loop gives $x^{-\NN}\mapsto x^{-\NN}e^{-2\pi i\NN}=x^{-\NN}M^{-1}$ and $\binom{u_{\rm g}}{u_{\rm d}}\mapsto M\binom{u_{\rm g}}{u_{\rm d}}$ by the above construction. Thus $\Phi$ does not contain logarithms, i.e.\ it gives the pure-power parts of the solutions. All the logarithmic terms are therefore generated by $x^{\NN}=\sum_j\NN^j\frac{1}{j!}\log^jx$.
	
	\item[]\textit{(iii) The entries of $\NN$:} The coefficient of $\log x$ in $u_{\rm g}$ is $\NN_{11}\Phi_{\rm g}+\NN_{12}\Phi_{\rm d}$. Because of the pure-power normalization that we adopt, $\Phi_{\rm g}$ has coefficient $1$ at $x^{-\ell-1}$ and $0$ at $x^{\ell}$, and $\Phi_{\rm d}$ has coefficient $0$ at $x^{-\ell-1}$ and $1$ at $x^{\ell}$, exactly to all orders in $\omega^2$. Reading the coefficients of $x^{-\ell-1}\log x$ and $x^{\ell}\log x$ in $u_{\rm g}$ therefore gives $\NN_{11}=C$ and $\NN_{12}=-A$ as defined in \eqref{eq:ADBdef}. Similarly $\NN_{21}=B$ and $\NN_{22}=\bar C$ from $u_{\rm d}$.
	
	\item[]\textit{(iv) $\NN$ is traceless:} The Wronskian $W=u_{\rm g}u_{\rm d}'-u_{\rm g}'u_{\rm d}$ obeys $W'=-PW$ with $P(x):=\frac{p(x)}{x}$. Since $p$ is analytic, $W$ is single-valued upon analytic continuation, i.e.\ $W\mapsto W$ around a closed loop.\footnote{Notice that $p_0$ is an integer because by assumption the Frobenius exponents are integers. Therefore $W\propto x^{-p_0}\times$(analytic) is single-valued.} On the other hand, the monodromy $\binom{u_{\rm g}}{u_{\rm d}}\mapsto M\binom{u_{\rm g}}{u_{\rm d}}$ gives $W\mapsto \det(M)\,W=e^{2\pi i\,{\rm tr}\NN}W$. Therefore ${\rm tr}\,\NN$ must be an integer. However, since $\NN(\omega^2)$ is continuous in $\omega^2$ and ${\rm tr}\,\NN(0)=0$, we conclude that ${\rm tr}\,\NN=0$. This proves that $\bar C=-C$.
	
	The final statement about the square of $\NN$ then follows immediately from the Cayley-Hamilton theorem: any traceless $2\times2$ matrix satisfies $\NN^2=-\det\NN\,\mathbf 1=(C^2-AB)\mathbf 1$.

\end{itemize}

\subsection{Renormalized angular momentum}

The above theorem has an interesting application in relation to the notion of `renormalized angular momentum' that one encounters in the MST formalism~\cite{Mano:1996mf,Mano:1996vt,Sasaki:2003xr}. The main observation is that the monodromy $M=e^{2\pi i \NN}$ in fact captures the `true' monodromy of the system, in the sense that the solutions near $x=0$ (i.e.\ not in the near-zone) behave as
\beq \label{eq:MST1}
u_{\rm g}\sim x^{-(\ell-\delta)-1} \,,\qquad u_{\rm d}\sim x^{\ell-\delta} \,.
\eeq
This follows directly from the normal form decomposition and the fact that $\NN$ has eigenvalues $\pm\delta$ ($\NN$ is traceless and its determinant is $-\delta^2$). In other words, the resummation of the logarithms achieved by the monodromy yields the correct leading powers of the exact solution. Although \eqref{eq:theorem} was derived in the near-zone, the monodromy is independent of the loop one takes around the singularity, and is thus valid everywhere up to the location of the nearest singularity. This is to be contrasted with the `formal' monodromy that one computes when studying the solutions in the vicinity of $x=0$ using the eikonal or WKB approximation, in which case one finds $u\sim e^{\pm i\omega}x^{\pm i\omega}$ (up to some constant factors in the exponents depending on the precise form of the equation). Naively reading off the monodromy as ${\rm diag}(e^{-2\pi\omega},e^{2\pi\omega})$ yields an incorrect result, a manifestation of the Stokes phenomenon~\cite{Castro:2013lba}.

MST considered the construction of solutions of the Regge-Wheeler and Teukolsky equations in terms of Coulomb functions, finding the asymptotic behaviors $u_{\rm g}\sim x^{-\nu_{\rm MST}-1}$ and $u_{\rm d}\sim x^{\nu_{\rm MST}}$, where the parameter $\nu_{\rm MST}$ `renormalizes' the angular momentum number $\ell$ in the presence of nonzero frequency~\cite{Fujita:2004rb,Fujita:2005kng},
\beq
\nu_{\rm MST}=\ell +\nu_2\omega^2+\nu_4\omega^4+\cdots \,.
\eeq
The interpretation of the renormalized angular momentum in terms of monodromy therefore allows us to conclude that, as a formal power series in $\omega^2$,
\beq
\nu_{\rm MST}=\ell -\delta \,.
\eeq
We should emphasize that this identification is not new~\cite{Castro:2013kea,Castro:2013lba} (see also~\cite{Aggarwal:2019iay,Chanson:2020hly,DiRusso:2024hmd,Aminov:2024mul,Pradhan:2026czt} for other applications of black hole monodromy). What is perhaps underappreciated is that $\delta$, and hence $\nu_{\rm MST}$, can be efficiently computed through an explicit algebraic recurrence relation. Moreover, and importantly, this applies to any metric background, even though the MST and monodromy constructions have so far only been applied to some particular black hole spacetimes.

We will gain a sense of the advantages of our method when we study some applications in Sec.\ \ref{sec:black holes}. Here we make a simple observation from the general structure of the three logarithmic coefficients $A$, $B$ and $C$. Explicitly expanding the determinant,
\begin{equation}
	\delta^2=C^2-AB=\gamma_1^2\omega^4+2\gamma_1\gamma_2\omega^6+\cdots-\alpha_0\beta_{\ell+1}\omega^{2\ell+2}-\big(\alpha_1\beta_{\ell+1}+\alpha_0\beta_{\ell+2}\big)\omega^{2\ell+4}-\cdots \,.
	\label{eq:delta square explicit}
\end{equation}
We see that, for $\ell\geq 2$, $\delta^2$ is approximated by $C^2$, i.e.\
\beq
\nu_{\rm MST}=\ell -C+\mathcal{O}(\omega^{2\ell}) \,,
\eeq
where the correction $\mathcal{O}(\omega^{2\ell})$ is present only if $\alpha_0\neq0$, otherwise it is $\mathcal{O}(\omega^{2\ell+2})$. In particular, for gravitational perturbations with $\alpha_0=0$, $\nu_2=-\gamma_1$ and $\nu_4=-\gamma_2$ for all $\ell\geq2$.

We have been cavalier about signs. For $\ell\geq2$, we have $\delta^2\simeq \gamma_1^2\omega^4$ at leading order, and we can define $\delta$, and hence $\nu_{\rm MST}$, such that $\delta\simeq \gamma_1\omega^2$, whatever the sign of $\gamma_1$. This is pure convention, since physically $\nu_{\rm MST}$ is defined modulo $\nu_{\rm MST}\to-\nu_{\rm MST}-1$ and integer shifts. All this remains valid for $\ell=0,1$, only that in these cases $\delta^2$ is not necessarily positive, even perturbatively in $\omega^2$. That is, for $\ell=0$ we define $\delta$ such that $\delta\simeq \sqrt{-\alpha_0\beta_1}\,\omega$ if $\alpha_0^{(\ell=0)}\neq0$ ($\beta_1=2$, cf.\ \eqref{eq:beta1 result}) at leading order, or $\delta\simeq\sqrt{\gamma_1^2-\alpha_1\beta_1}\,\omega^2$ if $\alpha_0^{(\ell=0)}=0$.\footnote{For a minimally coupled scalar field on any static and spherically symmetric background, $\alpha_0^{(\ell=0)}=0$ holds~\cite{Garcia-Saenz:2022wsl}. Presumably this remains true upon including non-minimal couplings, although we are not aware of a full proof.} In either case, the result is not guaranteed to be real. Similarly for $\ell=1$, $\delta\simeq\sqrt{\gamma_1^2-\alpha_0\beta_2}\,\omega^2$.

The fact that $\nu_{\rm MST}$ can become imaginary is a known fact in GR~\cite{Fujita:2004rb,Fujita:2005kng}. However, for Schwarzschild and Kerr this only happens above a critical value of the frequency, i.e.\ $\delta^2$ is real at leading order in $\omega^2$; see Sec.\ \ref{sec:black holes}. Our results instead show that, even at lowest order for $\ell=0,1$ perturbations, the condition of stability may translate into bounds on the black hole parameters. We will see this concretely in Sec.\ \ref{sec:black holes}.

\subsection{(Non-)universality}

A nice property of the monodromy generator is that it transforms covariantly under changes of basis. That is, if we let $\binom{\tilde{u}_{\rm g}}{\tilde{u}_{\rm d}}:=S\binom{u_{\rm g}}{u_{\rm d}}$ for a constant, invertible matrix $S(\omega^2)$, then $\binom{\tilde{u}_{\rm g}}{\tilde{u}_{\rm d}}=x^{\tilde\NN}S\Phi$, with $\tilde\NN:= S\NN S^{-1}$. Physically, it is reasonable to insist that $\tilde{u}_{\rm d}$ should still decay as $x^{\ell}$ for $\omega=0$, i.e.\ we demand that $S$ should not add $\tilde{u}_{\rm g}$ to $\tilde{u}_{\rm d}$. The most general such $S$ is (here the primes are just labels)
\begin{equation}
	S=\begin{pmatrix}c&c''\\0&c'\end{pmatrix}:\qquad
	C\to C+\frac{c''}{c}B \,,\qquad A\to\frac{c}{c'}\Big(A+2\frac{c''}{c}C+\frac{c''^2}{c^2}B\Big) \,,\qquad B\to\frac{c'}{c}B \,.
	\label{eq:covariance}
\end{equation}
This contains, in particular, the usual ambiguity of adding a decaying solution to the growing one, and we see that the logarithmic coefficients are generally affected~\cite{Katagiri:2024wbg,Kobayashi:2025vgl}. Nevertheless, the covariance property of $\NN$ allows us to understand this scheme-dependence, with the following observations:
\begin{itemize}
	\item The determinant $\delta^2=C^2-AB$ is of course invariant, justifying that $\nu_{\rm MST}$ is a robust observable.
	
	\item If we take $c_0=c'_0$ at level 0 ($\omega=0$), which amounts to simply normalizing the solutions, then $\alpha_0$ is invariant. This is the well-known universality of the static logarithmic Love number.
	
	\item Suppose $S_0=\mathbf{1}$ at level 0. Then, at level 1, $\alpha_1\to \alpha_1+(c_1-c'_1)\alpha_0$, so $\alpha_1$ is invariant if and only if $\alpha_0=0$.
	
	\item Under a static shift with $c''_0\neq0$ (the usual source/response ambiguity) then, at level 1, $\alpha_1\to \alpha_1+2c''_0 \gamma_1+(c''_0)^2\beta_1 \delta_{\ell,0}$ and $\gamma_1\to \gamma_1+c''_0\beta_1 \delta_{\ell,0}$ (here $\delta_{n,m}$ is the Kronecker delta; recall that $\beta_1$ is present only for $\ell=0$).
\end{itemize}

We will comment further on these ambiguities in Sec.\ \ref{sec:renormalization} when we discuss the renormalization group equation. The simple lesson here is that the logarithmic coefficients are generally scheme-dependent, and therefore results should be interpreted keeping this in mind. The pure-power scheme that we adopt in this paper is particularly nice because it agrees with the familiar trick of analytically continuing $\ell$ to non-integer values in the static problem~\cite{LeTiec:2020spy,LeTiec:2020bos,Creci:2021rkz}, as is evident from \eqref{eq:MST1}. However it disagrees with schemes commonly used in the world-line EFT literature; in Sec.\ \ref{sec:black holes} we will make an explicit comparison in terms of the above dictionary.

What about more general redefinitions? The static Love number ambiguity is often discussed in terms of reparametrizations of the radial variable. So let us consider $x=\psi(\tilde x)=\tilde x+c_1\tilde x^2+c_2\tilde x^3+\cdots$ (where the $c$'s here should not be confused with the ones above). This preserves the \textit{form} of the original ODE, Eq.\ \eqref{eq:general ode}, with the same indicial polynomial. So everything goes through, and we can define a basis of solutions $\tilde U$ in the pure-power scheme, with an associated monodromy generator. In fact, the latter remains unchanged, simply because the singularity $x=0=\tilde x$ is the same. The only difference shows up in the pure-power part $\Phi$: $U=x^{\NN}\Phi(x)=\tilde x^{\NN}\tilde{\Phi}(\tilde x)$, with $\tilde{\Phi}(\tilde x):=e^{\NN\log(\psi/\tilde x)}\Phi(\psi(\tilde x))$, which we note is analytic, as claimed. The conclusion is then that reparametrizations of the radial variable are equivalent to the changes of basis already discussed: $U$ is a solution for the transformed equation (obviously, since we have only redefined $x$), however it will not in general be a basis in the pure-power scheme.

A third type of field redefinition of interest in the black hole perturbation literature is the Darboux transformation, with the Chandrasekhar symmetry relating the Regge-Wheeler and Zerilli equations as the classic example~\cite{Chandrasekhar:1975nkd,Chandrasekhar:1985kt} (see also~\cite{Hui:2020xxx,Rosen:2020crj} for modern applications). It is defined by the field redefinition $u=\mathcal{A}(x)\tilde u+\mathcal{B}(x)\tilde u'$, and at least in the black hole context one focuses on the case where $\mathcal{A}$ and $\mathcal{B}$ are analytic at $x=0$. This latter property means that the monodromy structure is again preserved, even if the form of the equation does in general change; in more plain language, a Darboux transformation commutes with analytic continuation in the vicinity of $x=0$, and this is sufficient to infer that the monodromy generator will transform covariantly, with the consequences discussed above.


\section{Dynamical running} \label{sec:renormalization}

\subsection{Renormalization group flow}

Recall that we have defined the Love number $\mathcal{K}(\omega)$ as in Eq.\ \eqref{eq:uphys}, i.e.\ as the ratio between growing and decaying solutions: $u_{\rm phys}=u_{\rm g}+\mathcal{K}(\omega)u_{\rm d}$. While both $u_{\rm g}$ and $u_{\rm d}$ are even in $\omega$, the horizon boundary condition breaks this property, so $\mathcal{K}(\omega)$ generally includes all powers, $\mathcal{K}(\omega)=\mathcal{K}_0+i\mathcal{K}_1\omega+\mathcal{K}_2\omega^2+i\mathcal{K}_3\omega^3+\cdots$, where we have made explicit that the odd powers have imaginary coefficients, interpreted as measures of dissipation due to black hole accretion.

The decomposition $u_{\rm phys}=u_{\rm g}+\mathcal{K}(\omega)u_{\rm d}$ is ambiguous because $u_{\rm g}$ contains terms proportional to $u_{\rm d}$, in particular $\left[A(\omega^2)\log(r/r_g)+\cdots\right]u_{\rm d}$ (where we reinstate the Schwarzschild radius here for clarity), as the normal form decomposition makes clear. An observer at coordinate distance $r$ who measures the tidal and response fields in terms of the powers $r^{\ell+1}$ and $r^{-\ell}$ would measure a Love number which in fact runs with $r$. We can rephrase this in the language of the renormalization group, by noting that the choice of using $r_g$ as a reference scale in the logarithm is of course arbitrary. Replacing $\log(r/r_g)$ with $\log(\mu r)$ gives the basis of solutions
\beq
U(r;\mu)=\begin{pmatrix}u_{\rm g}(r;\mu) \\ u_{\rm d}(r;\mu) \end{pmatrix}=e^{-\NN \log\mu}U(r) \,,
\eeq
where $U(r)$ is our old basis defined in terms of $r_g$ (set again to 1 here), and $\mu$ is an arbitrary scale. Notice that $U(r;\mu)$ is still a basis in the pure-power scheme, but normalized with respect to powers of $\mu r$. In this new basis the physical solution has the form
\beq \label{eq:uphys in terms of mu}
u_{\rm phys}=\mathcal{Z}(\omega;\mu)\left[u_{\rm g}(r;\mu)+\mathcal{K}(\omega;\mu)u_{\rm d}(r;\mu)\right] \,,
\eeq
where $\mathcal{K}(\omega;\mu)$ defines the Love number at the scale $\mu$, and $\mathcal{Z}(\mu)$ is some function that can we interpret as a `wave-function renormalization' of the source. All this is in precise analogy with the well-known results of the world-line EFT, as reviewed in the Introduction, where logarithmic divergences generically arise. In the static limit, or order by order in $\omega$, these are logarithms of $r$ (in position space, the only infrared scale associated to a massless bulk field), and upon matching are thus in one-to-one correspondence with the logarithms discussed here. Our formalism clarifies how this works at every order in $\omega$ and its applicability to arbitrary background metrics.

Let us write $u_{\rm phys}$ as $(1\;,\; \mathcal{K}(\omega))U(r)=(1\;,\; \mathcal{K}(\omega))e^{\NN\log\mu}U(r;\mu)$, where $(1\;,\; \mathcal{K}(\omega))$ denotes a row vector. Comparing this with \eqref{eq:uphys in terms of mu} we infer
\beq\bal \label{eq:flow solution}
\mathcal{K}(\omega;\mu)&=\frac{\cosh(\delta\log\mu)\mathcal{K}-\delta^{-1}\sinh(\delta\log\mu)(A+C\mathcal{K})}{\cosh(\delta\log\mu)+\delta^{-1}\sinh(\delta\log\mu)(C+B\mathcal{K})} \,,\\
\mathcal{Z}(\omega;\mu)&=\cosh(\delta\log\mu)+\delta^{-1}\sinh(\delta\log\mu)(C+B\mathcal{K}) \,,
\eal\eeq
using that $e^{\NN\log\mu}=\cosh(\delta\log\mu)\mathbf{1}+\delta^{-1}\sinh(\delta\log\mu)\NN$ and $\NN^2=\delta^2\mathbf{1}$. Thus $\mathcal{K}(\omega;\mu)$ is related to $\mathcal{K}(\omega)$ via a M\"obius transformation. Differentiating,
\beq\boxed{\bal \label{eq:flow}
\frac{\partial\mathcal{K}(\omega;\mu)}{\partial\log\mu}&=-A(\omega^2)-2C(\omega^2)\mathcal{K}(\omega;\mu)-B(\omega^2)\mathcal{K}(\omega;\mu)^2 \\
\frac{\partial\log\mathcal{Z}(\omega;\mu)}{\partial\log\mu}&=C(\omega^2)+B(\omega^2)\mathcal{K}(\omega;\mu) \\
\eal}\eeq
We emphasize that this is exact. The fact that the renormalization group flow depends only on the three logarithmic coefficients $A$, $B$ and $C$ is again a direct result of the resummation of all logarithms effected by the monodromy. The quadratic form of the flow equation, only visible in the dynamical case, was understood only recently from direct calculations done for Schwarzschild~\cite{Caron-Huot:2025tlq,Chang:2026eti,Solon:2026ubm} (see also~\cite{Aminov:2024mul}). Our method clarifies that this is a direct consequence of the above M\"obius structure and thus applies to any type of tidal perturbation on any black hole background. Moreover, and for the same reason, the quadratic structure is preserved under the scheme ambiguities discussed in the previous section: covariant transformations of $\NN$ lead to the same M\"obius form, only with different $A$, $B$ and $C$, whose scheme dependence is known.

Notice that the amplitude of the tidal field runs with $C$, while the amplitude of the response, $\mathcal{Z}\mathcal{K}$, runs with $-C$,
\beq\bal
\frac{\partial\mathcal{Z}(\omega;\mu)}{\partial\log\mu}&=\mathcal{Z}(\omega;\mu)\left(C(\omega^2)+B(\omega^2)\mathcal{K}(\omega;\mu)\right) \,,\\ \frac{\partial(\mathcal{Z}(\omega;\mu)\mathcal{K}(\omega;\mu))}{\partial\log\mu}&=-\mathcal{Z}(\omega;\mu)\left(A(\omega^2)+C(\omega^2)\mathcal{K}(\omega;\mu)\right) \,.
\eal\eeq
Thus $C$ has an interpretation as anomalous dimension of the field, exactly the same but with opposite signs for the growing and decaying parts. This is another direct consequence of the general theorem, specifically the statement that $\NN$ is traceless. This is also manifest in the anomalous exponents of Eq.\ \eqref{eq:MST1}, remembering that $\delta\simeq C$ at leading order (when the corrections from $B$ are negligible). The ratio of the two scales as $x^{-2\delta}$, in agreement with \eqref{eq:flow}.

The term $B\mathcal{K}^2$, which we recall enters at order $\omega^{2\ell+2}$, can be interpreted as a tail-like effect, to draw the language of the post-Newtonian literature~\cite{Blanchet:2002av,Blanchet:2006zz,Blanchet:2013haa,Galley:2015kus}. $B$ is the coefficient of $u_{\rm g}\log x$ inside $u_{\rm d}$, i.e.\ a modification to the response by the running of the source. This also sheds light into why the leading coefficient $\beta_{2\ell+1}$ is universal, the same for any metric and tidal field (we prove this below in Sec.\ \ref{sec:black holes}; the case $\ell=0$ is in Eq.\ \eqref{eq:beta1 result}): it is the tail effect of the leading part of the gravitational potential, i.e.\ the Newtonian term, hence completely fixed by the mass of the object.

We should remember that everything is assumed to be perturbative in $\omega$. This implies that the flow equation \eqref{eq:flow} must be `solved' as a hierarchy of ODEs. Explicitly, assuming for simplicity $\ell\geq 2$ (i.e.\ $B$ starts at order $\geq6$), we have
\beq\bal
\frac{\d\mathcal{K}_0}{\d\log\mu}&=-\alpha_0 \,,\\
\frac{\d\mathcal{K}_1}{\d\log\mu}&=0 \,,\\
\frac{\d\mathcal{K}_2}{\d\log\mu}&=-\alpha_1-2\gamma_1\mathcal{K}_0 \,,\\
\frac{\d\mathcal{K}_3}{\d\log\mu}&=-2\gamma_1\mathcal{K}_1 \,,\\
\frac{\d\mathcal{K}_4}{\d\log\mu}&=-\alpha_2-2\gamma_1\mathcal{K}_2-2\gamma_2\mathcal{K}_0 \,,\\
\eal\eeq
for the first few orders. Obviously one can only solve these given the `initial condition' $\mathcal{K}(\omega)$, as of course we knew already since our starting point, Eq.\ \eqref{eq:flow solution}, was nothing but the exact solution.

As a last comment in relation to \eqref{eq:flow solution}, we recall from the previous section that $\delta$ may be real or purely imaginary. Notice that $\delta^2=C^2-AB$ is, up to a factor, the discriminant of the quadratic form in \eqref{eq:flow}. Thus, although $\mathcal{K}(\omega;\mu)$ is real in either case, as it should, the fixed points of \eqref{eq:flow} could merge and then become complex in a scenario where $\delta^2$ changes from positive to negative. This effect has been discussed in the context of strongly-coupled systems as a mechanism to describe conformal to non-conformal phase transitions~\cite{Kaplan:2009kr}. The caveat here is that the fixed points lie at $|\log\mu|\sim \delta^{-1}\sim \omega^{-2}$, which for $\mu\sim\omega$ means $\omega^2|\log\omega|\sim 1$, outside the domain of validity of the perturbative expansion at that scale (see below). It would nevertheless be interesting to investigate this aspect further, particularly in the context of the world-line EFT.

\subsection{The energy of the `experiment'}

To calculate observables in perturbation theory, one would like to choose the renormalization scale $\mu$ to match the typical energy of the `experiment', which in the setting of dynamical tidal deformations is naturally identified with the frequency $\omega$. This appears as problematic in the near-zone expansion, where the choice $\mu\sim\omega$ generates $\log(\omega r)$, i.e.\ large logarithms. The point is that such large logarithms are not an issue, for the simple reason that we have their full resummation thanks to \eqref{eq:theorem}. The choice $\mu\sim\omega$ is valid, but is best interpreted as `natural' in the sense of an observer located in the far-zone who organizes the perturbative expansion in terms of $(\omega r)^{-1}\ll1$. Indeed, the trivial identity $\log(\mu r)=\log(\omega r)+\log(\mu/\omega)$ shows that the logarithms of the near-zone expansion are the same logarithms of the far-zone expansion, upon running to the scale $\mu\sim\omega$. This \textit{still} leaves us with (exactly the same) large logarithms though, which is just a feature of the perturbative expansion that one must take into account.

As far as the Love number is concerned, we can be more quantitative thanks to the above results. For GR black holes we have $\mathcal{K}_0=0=\alpha_0$ so that
\beq
\mathcal{K}(\omega;\omega)=i\mathcal{K}_1\omega+\left(\mathcal{K}_2-\alpha_1\log\omega\right)\omega^2+\cdots \,.
\eeq
The logarithm here may well dominate the second term, but it is safely subleading relative to the leading order dissipative term. Nevertheless, the relative correction of order $\omega|\log\omega|$ is at the few percent level for $\omega=\mathcal{O}(10^{-2})$, for instance. Interestingly, the situation is markedly different for black holes beyond GR, which generically have $\mathcal{K}_0\neq0\neq\alpha_0$. This now gives $\mathcal{K}(\omega;\omega)=\mathcal{K}_0-\alpha_0\log\omega+\cdots$ and thus the logarithm gives an obviously important contribution.


\section{Applications to black holes} \label{sec:black holes}

Everything discussed so far applies to any tidal perturbation field on any static and spherically symmetric background, subject only to the assumptions we have made regarding the structure of the ODE \eqref{eq:general ode}. It is applicable in particular to any asymptotically flat black hole metric, which we write as
\beq
\d s^2=-f(r)\d t^2+\frac{\d r^2}{g(r)}+r^2\d\Omega^2 \,,
\eeq
with the mild condition that $f$ and $g$ are analytic at $x=0$, i.e.\
\beq \label{eq:general metric ansatz}
f=1-x\bigg(1+\sum_{n\geq1}f_nx^n\bigg) \,,\qquad g=1-x\bigg(1+\sum_{n\geq1}g_nx^n\bigg) \,.
\eeq
This includes essentially all metrics of interest in modified gravity, including regular black holes, as well as EFT-corrected metrics provided the above series are understood as perturbative.

\subsection{General results}

Some general results can be inferred before specifying the type of tidal perturbation or black hole background. One only needs to assume that the perturbation obeys a Schr\"odinger-type equation in terms of the tortoise coordinate $\d r^{*}=(fg)^{-1/2}\d r$, i.e.\
\beq \label{eq:schrodinger}
\frac{\d^2u}{\d r^{*2}}+\left(\omega^2-V\right)u=0 \,.
\eeq
Transforming to our canonical form \eqref{eq:general ode} yields
\beq \label{eq:general p q R}
p=2+\frac{x}{2}\left(\frac{f'}{f}+\frac{g'}{g}\right) \,,\qquad q=-\frac{V}{x^2fg} \,,\qquad \mathcal{R}=\frac{1}{x^2fg} \,.
\eeq
Thus $p$ and $\mathcal{R}$ are fully specified in terms of the metric functions; all dependence on $\ell$ and on the type of perturbation is encapsulated by the potential $V$ and hence by $q$.

\subsubsection{Residue formulae}

The above explicit expression for $\mathcal{R}$ means that we essentially know the form of the source term at level 1, $\hat{L}_0u^{(2)}=-\mathcal{R}u^{(0)}$. Although we cannot know the exact solution $u^{(0)}$ without solving the level-0 problem, their Wronskian is explicitly known thanks to Abel's identity,
\beq
W^{(0)}=u_{\rm g}^{(0)}u_{\rm d}^{(0)\prime}-u_{\rm g}^{(0)\prime}u_{\rm d}^{(0)}\propto \exp\left[-\int \frac{p}{x}\,\d x\right]\qquad \Rightarrow\qquad W^{(0)}=\frac{2\ell+1}{x^2\sqrt{fg}} \,,
\eeq
where the proportionality constant is fixed by our choice of unit pure-power normalization of the asymptotic solutions, $u_{\rm g}^{(0)}\sim x^{-\ell-1}$ and $u_{\rm d}^{(0)}\sim x^{\ell}$.

Given $W^{(0)}$ we can then write down the solutions at level 1, in terms of $u_{\rm g}^{(0)}$ and $u_{\rm d}^{(0)}$, via the variation of parameters formula. Again, without knowing the latter, not much can be said about the details of these solutions. But it does have a nice application in relation to the logarithmic coefficients $\alpha_1$ and $\gamma_1$ (cf.\ \eqref{eq:ADBdef}), stated in the following proposition.

\textbf{Proposition.} Suppose that $\alpha_0=0$ for a given multipole $\ell$. Then $\alpha_1$ and $\gamma_1$ are determined in terms of residues at $x=0$,
\beq \label{eq:residue1}
\alpha_1=\frac{1}{2\ell+1}\mathop{\rm Res}_{x=0}\,\frac{u_{\rm g}^{(0)2}}{x^2\sqrt{fg}} \,,\qquad \gamma_1=\frac{1}{2\ell+1}\mathop{\rm Res}_{x=0}\,\frac{u_{\rm g}^{(0)}u_{\rm d}^{(0)}}{x^2\sqrt{fg}} \,.
\eeq

The proof follows from the variation of parameters formula. Since $\alpha_0=0$, neither $u^{(0)}$ contains logarithms and the above functions are just Laurent series. They produce logarithms in the level-1 solutions through their $1/x$ terms, i.e.\ their residues. In fact, only $u_{\rm g}^{(0)}\subset u^{(0)}$ contains poles, thus\footnote{Here the `source' is $S=-\mathcal{R}u^{(0)}/x^2$ because the variation of parameters formula uses the form $u''+Pu'+Q=S$, which is our canonical ODE \eqref{eq:general ode} divided by $x^2$.}
\beq
u^{(1)}=-u_{\rm g}^{(0)}\int^x\frac{u_{\rm d}^{(0)}(-\mathcal{R}u_{\rm g}^{(0)}/x^2)}{W^{(0)}}+u_{\rm d}^{(0)}\int^x\frac{u_{\rm g}^{(0)}(-\mathcal{R}u_{\rm g}^{(0)}/x^2)}{W^{(0)}}+({\rm analytic}) \,,
\eeq
Reading off the coefficients of $x^{\ell}$ and $x^{-\ell-1}$ then yields \eqref{eq:residue1}.

The proposition is in general not terribly useful unless one can compute the series coefficients of $u_{\rm g}^{(0)}$ and $u_{\rm d}^{(0)}$, which basically takes us back to the Frobenius analysis of Sec.\ \ref{sec:monodromy}. Its advantage lies in the residue theorem applied to meromorphic functions, i.e.\ the statement that the sum of all residues, including $x=\infty$, is zero.

To make this concrete, let us assume $g=f$ and that $u_{\rm g}^{(0)}$ is a Laurent polynomial in $x$, i.e.\ its Frobenius series terminates. Suppose further that the residue at infinity of $u_{\rm g}^{(0)2}/(x^2f)$ vanishes. These are of course strong assumptions, but actually encompass the situation of GR black holes, as well as monopole scalar perturbations for general black holes, as we recollect below. Then the residue theorem implies
\beq \label{eq:residue2}
\alpha_1=-\frac{1}{2\ell+1}\sum_{f(\xi_i)=0}\mathop{\rm Res}_{x=\xi_i}\,\frac{u_{\rm g}^{(0)2}}{x^2f} \,,
\eeq
i.e.\ a sum over the residues at the black hole horizons.

\subsubsection{Universal leading tail term}

In Sec.\ \ref{sec:monodromy} we found that $\beta_{\ell+1}$, i.e.\ the leading logarithmic term in the `tail' term $B$ (cf.\ \eqref{eq:ADBdef}), is universal for $\ell=0$, Eq.\ \eqref{eq:beta1 result}. This generalizes to all multipoles $\ell$ modulo an extra assumption on the form of the potential $V$ in \eqref{eq:schrodinger}. Namely, we assume that $V$ is such that
\beq \label{eq:q0}
q_0=-\ell(\ell+1) \,.
\eeq
This is nothing but the centrifugal contribution to the potential and therefore the assumption is expected to hold very generally. From \eqref{eq:general p q R} we also have 
\beq \label{eq:p0 p1 w0 w1}
p_0=2 \,,\qquad p_1=-1 \,,\qquad w_0=1 \,,\qquad w_1=2 \,,
\eeq
for any Schr\"odinger-type equation. These data are enough to determine $\beta_{\ell+1}$ for all $\ell$, with the result
\beq \label{eq:beta all ell}
\boxed{\beta_{\ell+1}=\frac{2^{2\ell+1}(\ell!)^2}{(2\ell+1)!(2\ell)!}
}
\eeq
Notice that the fact that $\beta_{\ell+1}$ does not depend on the (non-universal) coefficient $q_1$ is non-trivial. The proof is given in Appendix \ref{app:beta universal}.

\subsubsection{Almost universal leading anomalous dimension term}

The leading term $\gamma_1$ in the anomalous dimension coefficient $C$ can also be expressed in closed form using the result \eqref{eq:delta1 recursion}. It is however not completely universal as it is sensitive to the first asymptotic corrections in the metric, $f_1$ and $g_1$, as well as to the first two non-universal terms in the potential.

We set $g=f$ here for simplicity. Then, in addition to \eqref{eq:q0} and \eqref{eq:p0 p1 w0 w1}, here we need
\beq
p_2=-1-2f_1 \,,\quad w_2=3+2f_1 \,,\quad q_1=-\ell(\ell+1)-U_1 \,,\quad q_2=-\ell(\ell+1)(1+f_1)-U_1-U_2 \,,
\eeq
where $U_n$ denotes the $n$-th Taylor coefficient of $U:=\frac{V}{x^2f}=\ell(\ell+1)+U_1x+U_2x^2+\cdots$. A direct evaluation of formula \eqref{eq:delta1 recursion} yields
\beq \label{eq:general gamma1}
\boxed{\gamma_1=\frac{15\ell(\ell+1)-5-\frac{3}{\ell(\ell+1)}+12\ell(\ell+1)f_1-6U_1-4U_2+\frac{3U_1^2}{\ell(\ell+1)}}{2(2\ell-1)(2\ell+1)(2\ell+3)}
}
\eeq

\subsection{Regge-Wheeler equation}

The Schwarzschild black hole spacetime provides a good test bed for our methods. The metric functions are \eqref{eq:general metric ansatz} with all $f_n,g_n=0$. We focus on the Regge-Wheeler equation, relevant for scalar and odd-parity vector and tensor perturbations. It conforms to the Schr\"odinger-type form \eqref{eq:schrodinger} with
\beq
V=x^2(1-x)\left[\ell(\ell+1)+(1-s^2)x\right]\qquad \Rightarrow\qquad q=-\frac{\ell(\ell+1)+(1-s^2)x}{1-x} \,,
\eeq
where $s=0,1,2$ respectively for scalar, vector and tensor perturbations.

Mechanically running the master recurrence formula we find the results in Table \ref{tab:bh-schw} for the leading logarithmic coefficients. The vanishing of $\alpha_0$ is of course the well-known `no-Love' theorem for GR black holes~\cite{}. The tail terms $\beta_{\ell+1}$ are in full agreement with \eqref{eq:beta all ell}, while the results for the anomalous dimension term $\gamma_1$ also verify the general formula \eqref{eq:general gamma1} upon setting $f_1=0$, $U_1=1-s^2$ and $U_2=0$.
\begin{table}[h]
	\centering
	\small
	\begin{tabular}{cccccccc}
		\toprule
		$s$ & $\ell$ & $\alpha_0$ & $\alpha_1$ & $\alpha_2$ & $\gamma_1$ & $\gamma_2$ & $\beta_{\ell+1}$\\
		\midrule
		0 & 0 & 0 & $1$ & $-\tfrac{23}{6}$ & $\tfrac{11}{6}$ & $-\tfrac{2291}{1080}$ & $2$\\
		0 & 1 & 0 & $\tfrac1{12}$ & $-\tfrac{217}{10800}$ & $\tfrac{19}{30}$ & $\tfrac{78037}{189000}$ & $\tfrac23$\\
		0 & 2 & 0 & $\tfrac1{180}$ & $-\tfrac{206}{23625}$ & $\tfrac{79}{210}$ & $\tfrac{708247}{9261000}$ & $\tfrac2{45}$\\
		0 & 3 & 0 & $\tfrac1{2800}$ & $-\tfrac{229793}{123480000}$ & $\tfrac{169}{630}$ & $\tfrac{74380421}{2750517000}$ & $\tfrac{2}{1575}$\\
		0 & 4 & 0 & $\tfrac1{44100}$ & $-\tfrac{19251641}{60511374000}$ & $\tfrac{289}{1386}$ & $\tfrac{435454879}{34612505928}$ & $\tfrac{2}{99225}$\\
		\midrule
		1 & 1 & 0 & $\tfrac13$ & $\tfrac{43}{540}$ & $\tfrac{47}{60}$ & $\tfrac{1148681}{1512000}$ & $\tfrac23$\\
		1 & 2 & 0 & $\tfrac1{80}$ & $-\tfrac{34259}{3528000}$ & $\tfrac{169}{420}$ & $\tfrac{6832249}{74088000}$ & $\tfrac2{45}$\\
		1 & 3 & 0 & $\tfrac1{1575}$ & $-\tfrac{28051}{11113200}$ & $\tfrac{233}{840}$ & $\tfrac{64437649}{2173248000}$ & $\tfrac{2}{1575}$\\
		\midrule
		2 & 2 & 0 & $\tfrac15$ & $\tfrac{21547}{126000}$ & $\tfrac{107}{210}$ & $\tfrac{1695233}{9261000}$ & $\tfrac2{45}$\\
		2 & 3 & 0 & $\tfrac1{252}$ & $-\tfrac{219319}{44452800}$ & $\tfrac{13}{42}$ & $\tfrac{10921}{271656}$ & $\tfrac{2}{1575}$\\
		2 & 4 & 0 & $\tfrac1{7056}$ & $-\tfrac{180382819}{193636396800}$ & $\tfrac{1571}{6930}$ & $\tfrac{68482418369}{4326563241000}$ & $\tfrac{2}{99225}$\\
		\bottomrule
	\end{tabular}
	\caption{Logarithmic coefficients for scalar ($s=0$), odd-parity vector ($s=1$) and odd-parity tensor ($s=2$) perturbations of a Schwarzschild black hole.}
	\label{tab:bh-schw}
\end{table}

The leading dynamical logarithmic coefficient $\alpha_1$ also has a closed-form formula,
\beq
\alpha_1=\frac{\big[(\ell-s)!(\ell+s)!\big]^2}{(2\ell)!(2\ell+1)!} \,.
\label{eq:a1-schw-s}
\eeq
This agrees with the results of Ref.~\cite{Combaluzier--Szteinsznaider:2025eoc} for $s=0$ and Ref.~\cite{Kobayashi:2025vgl} for $s=2$ (see also~\cite{Katagiri:2024wbg,Chakraborty:2025wvs}). It follows most easily from the residue formula \eqref{eq:residue2}, using the known exact solutions of the static Regge-Wheeler equation (see e.g.~\cite{Hui:2020xxx}). In our normalization,\footnote{For $s=0$ the result may be expressed more neatly in terms of Legendre polynomials, in a way that makes manifest the value at the horizon $x=1$:
\beq
u_{\rm g}^{(0)}=\frac{(\ell!)^2}{(2\ell)!}\frac{1}{x}P_{\ell}\left(\frac{2}{x}-1\right) \,.
\eeq
}
\beq
u_{\rm g}^{(0)}=x^{-\ell-1}{}_2F_1\left(-\ell-s,-\ell+s,-2\ell;x\right) \,.
\eeq

Recall that for gravitational perturbations $\gamma_1$ and $\gamma_2$ coincide, up to a sign, with the leading terms in the MST renormalized angular momentum. For instance, for $\ell=2$, we compute
\beq
\nu_{\rm MST}^{(\ell=2)}=2-\frac{107}{210}\omega^2-\frac{1695233}{9261000}\omega^4+\mathcal{O}(\omega^6) \,.
\eeq
The $\mathcal{O}(\omega^2)$ term agrees with a known closed-form formula~\cite{Mano:1996vt}. We have furthermore checked that the $\mathcal{O}(\omega^4)$ term (for which no closed-form expression is known for all $\ell$) agrees with the numerical result obtained from the MST continued fraction formula.

Also instructive is to compare our results with those obtained in the EFT literature, specifically Ref.~\cite{Caron-Huot:2025tlq}, which computed dynamical Schwarzschild Love numbers up to $\mathcal{O}(\omega^7)$. In Table \ref{tab:eft comparison} we show the results of our calculation for the monopole $\ell=0$ ($s=0$) in the unit pure-power scheme for the three logarithms compared with the same numbers extracted from~\cite{Caron-Huot:2025tlq}. The latter were obtained in a $\overline{\rm MS}$ scheme, a fact that explains why the results are different. The discrepancy is in full agreement with the ambiguities discussed in Sec.\ \ref{sec:monodromy}: a redefinition of the basis of the form \eqref{eq:covariance} allows us to match the $\mathcal{O}(\omega^4)$ terms, with the choice $c_1-c_1'=-\tfrac{277}{72}$ and $c_1''=\tfrac{65}{24}$. We did not check the $\mathcal{O}(\omega^6)$ terms individually, however we find that the invariant
\beq
(\delta^{(\ell=0)})^2=\frac{49}{36}\omega^4+\frac{9449}{3240}\omega^6+\frac{71195267}{8164800}\omega^8+\mathcal{O}(\omega^{10}) \,,
\eeq
is indeed the same when computed in either scheme. It is worth stressing that this is a very non-trivial check considering that the two lines of Table \ref{tab:eft comparison} were obtained using different methods.
\begin{table}[h]
	\centering
\begin{tabular}{lccc}
	\toprule
	Scheme & $A$ & $C$ & $B$\\
	\midrule
	UPP & $\omega^2-\tfrac{23}{6}\omega^4+\tfrac{51199}{32400}\omega^6$ & $\tfrac{11}{6}\omega^2-\tfrac{2291}{1080}\omega^4+\tfrac{7687921}{1360800}\omega^6$ & $2\omega^2-\tfrac{109}{36}\omega^4+\tfrac{111991}{64800}\omega^6$ \\[2pt]
	$\overline{\rm MS}$ & $\omega^2+\tfrac94\omega^4+\tfrac{65051}{10800}\omega^6$ & $\tfrac{11}{6}\omega^2+\tfrac{3559}{1080}\omega^4+\tfrac{348893}{42525}\omega^6$ & $2\omega^2+\tfrac{14}{3}\omega^4+\tfrac{62701}{6480}\omega^6$\\
	\bottomrule
\end{tabular}
\caption{The three logarithms for monopole scalar perturbations ($\ell,s=0$) of a Schwarzschild black hole, expanded up to order $\mathcal{O}(\omega^6)$, as computed here in the unit pure-power scheme and in Ref.~\cite{Caron-Huot:2025tlq} in a $\overline{\rm MS}$ scheme.}
\label{tab:eft comparison}
\end{table}

\subsection{Moncrief equations}

Static Love numbers of the RN black hole have been studied in detail~\cite{Pereniguez:2021xcj,Rai:2024lho,Xia:2025zfp}, but the dynamical case appears to be unexplored. As a first study, we consider here odd-parity vector and tensor perturbations.

The RN metric corresponds to the form \eqref{eq:general metric ansatz} with $g=f$ and all $f_{n\geq2}=0$: $f=1-x-f_1x^2$. In terms of the horizon radii $r_+$ and $r_-$,
\beq
f_1=-r_+r_- \,,\qquad r_++r_-=r_g\equiv1 \,,\qquad -\tfrac{1}{4}\leq f_1\leq 0 \,,
\eeq
where the value $f_1=-1/4$ is the extremality bound. Axial electromagnetic and gravitational perturbations obey coupled equations, but can be decoupled in terms of suitable master variables and recast in Schr\"odinger form, resulting in the Moncrief equations~\cite{Moncrief:1974gw,Moncrief:1974ng,Moncrief:1975sb} (see also~\cite{Kodama:2003kk,Ishibashi:2011ws}). The potentials are given by
\beq
V_{\pm}=x^2f\left[\ell(\ell+1)-\mathcal{Q}_{\pm}x-4f_1x^2\right] \,,\qquad \mathcal{Q}_{\pm}:= \tfrac{3}{2}\pm\sqrt{\tfrac{9}{4}-4f_1(\ell-1)(\ell+2)} \,,\qquad \ell\geq2 \,.
\eeq
The $+/-$ equations reduce to the $s=2$ and $s=1$ Regge-Wheeler equations, respectively, upon setting $f_1=0$.

The static logarithmic Love number vanishes, $\alpha_0=0$ for all $\ell$. The static growing solution is again given by a Laurent polynomial, so the residue formula \eqref{eq:residue2} applies. We find\footnote{As usual the case $\ell=1$ must be dealt with separately. There is in fact a subtlety related to the scheme-dependence of $\alpha_1^{(\ell=1)}$, and the quantity we quote here is actually a scheme-invariant redefinition of it; see Appendix \ref{app:alpha1 RN} for the details. Subtleties aside, it turns out that the $\ell=1$ case can be cheekily obtained from the $\ell\geq2$ result by taking the limit $\ell\to1$ upon analytic continuation to real $\ell$.}
\beq\boxed{\bal
\alpha_1&=\frac{\big[(\ell-2)!(\ell+1)!\big]^2}{(2\ell)!(2\ell+1)!}\,(r_+^2+r_-^2)(r_+-r_-)^{2\ell-2}(\mathcal{Q}_{\pm}+\ell-1)^2 \,,\qquad \mbox{if}\quad\ell\geq2 \,,\\
\alpha_1&=\frac{(1+2f_1)(1+4f_1)^2}{3} \,,\qquad \mbox{if}\quad\ell=1 \,.
\label{eq:a1-RN}
\eal}\eeq
The derivation is in Appendix \ref{app:alpha1 RN}. We have also checked this result for the first few explicit values of $\ell$ using the master recurrence formula. Interestingly, we discover that $\alpha_1$ vanishes for an extremal RN black hole,
\beq
\boxed{\alpha_1^{\rm(extremal)}=0
}
\label{eq:a1-RNext}
\eeq
for all $\ell\geq1$. It is worth remarking that the above results are scheme-independent once the basis of static solutions is fixed, as explained in Sec.\ \ref{sec:monodromy}, because $\alpha_0=0$ for RN.

The leading anomalous dimension coefficient follows from the general formula \eqref{eq:general gamma1} with $U_1=-\mathcal{Q}_{\pm}$ and $U_2=-4f_1$. Evaluating at the extremality bound we obtain
\beq
\gamma_1^{\rm(extremal,+)}=\frac{6\ell^2+3}{8\ell^3-2\ell} \,,\qquad \gamma_1^{\rm(extremal,-)}=\frac{6\ell^2+12\ell+9}{8\ell^3+24\ell^2+22\ell+6} \,.
\eeq
Curiously, we find that the two are related,
\beq \label{eq:gamma1 RN extremal}
\boxed{\gamma_1^{\rm(extremal,+)}(\ell+1)=\gamma_1^{\rm(extremal,-)}(\ell)
}
\eeq
suggesting that there might be a symmetry relating the two equations in the extremal case mixing different multipoles. Moreover, we have checked that the above equality persists at level 2 for a few explicit values of $\ell$. We think these results deserve further investigation.\footnote{Presumably this coincidence is related to the fact that one can recast the two Moncrief equations into the same form through suitable field redefinitions~\cite{Rai:2024lho}. This by itself of course does not immediately prove \eqref{eq:gamma1 RN extremal}: first because the relation mixes different $\ell$, and second because one needs to take into account the change of scheme effected by the field redefinitions.}

\subsection{Klein-Gordon equation}

To discuss general metrics in an agnostic way, we must assume that the tidal field behaves as a probe which does not backreact on the spacetime. We focus here on a minimally coupled scalar field as the simplest case. The Klein-Gordon equation corresponds to the following choice of ODE coefficients:
\beq
p=2+\frac{x}{2}\left(\frac{f'}{f}+\frac{g'}{g}\right) \,,\qquad q=-\frac{\ell(\ell+1)}{g}+\frac{x}{2}\left(\frac{f'}{f}+\frac{g'}{g}\right) \,,\qquad \mathcal{R}=\frac{1}{x^2fg} \,.
\eeq

The generalization of \eqref{eq:general gamma1} to the case $g\neq f$ is easily deduced. For a probe scalar field the result depends only on the metric coefficients $f_1$ and $g_1$,
\beq
\gamma_1=\frac{15\ell(\ell+1)-11+4[2\ell(\ell+1)-1]f_1+4[\ell(\ell+1)-1]g_1}{2(2\ell-1)(2\ell+1)(2\ell+3)} \,.
\eeq
No simple closed-form formula is available for the leading dynamical coefficient $\alpha_1$, although it is a straightforward mechanical task to evaluate it for any given $\ell$ using the master recurrence. We can nevertheless establish general criteria for a given term in the metric to yield vanishing contributions to the logarithmic coefficients.

\textbf{Theorem.} Suppose $f=g=1-x(1+ax^N)$, $N\geq2$. Let the superscript $(\sigma)$ denote the $\mathcal{O}(a^\sigma)$ contribution to the logarithmic coefficients $\alpha_k$, $\beta_k$ and $\gamma_k$, at any level $k$ (understood as a formal power series in $a$). Then
\beq\bal
\alpha_k^{(\sigma)}&=0 \qquad \mbox{if}\quad 2\ell+1+2k<\sigma(N+1) \,,\\
\gamma_k^{(\sigma)}&=0 \qquad \mbox{if}\quad 2k<\sigma(N+1) \,,\\
\beta_k^{(\sigma)}&=0 \qquad \mbox{if}\quad 2k-2\ell-1<\sigma(N+1) \,.
\eal\eeq

See Appendix \ref{app:zero love theorem} for the proof. We have assumed $f=g$ for simplicity. It does in fact also apply for $f\neq g$, i.e.\ if we assume $f=1-x(1+c_1ax^N)$ and $g=1-x(1+c_2ax^N)$, where $c_{1,2}$ are numerical constants. Typically we expect $c_{1,2}$ to be both non-zero if the correction is present, for instance as in the case of EFT-corrected black hole metrics. However this is not necessary for the conclusion of the theorem to hold, i.e.\ either $c_1$ or $c_2$ may be taken to be zero. Notice also that the theorem applies if there are multiple deformations: if $f=g=1-x(1+ax^N+bx^M)$, then the conclusion follows \textit{separately} for $a$ and $b$.

The above result is interesting because it shows that the dynamical coefficients will always probe, at some level $k$, any deformation of the Schwarzschild or RN metric. Said otherwise, for any given $\sigma$, $N$ and $\ell$, there exists $k$ such that the coefficients will (generically) \textit{not} vanish. Take $N=5$ and $\sigma=1$ for example, which is the case for the leading corrections in the EFT of gravity from curvature cubed terms~\cite{Cardoso:2017cfl,DeLuca:2023mio,Noumi:2026shc}. The theorem then implies that $\alpha_1^{(1)}$ is generically non-zero when $\ell\geq2$, whereas the static Love number $\alpha_0^{(1)}$ is generically non-zero only when $\ell\geq3$, so that the quadrupole is sensitive to EFT corrections only at the dynamical level. The implication is that dynamical running Love numbers may potentially provide better targets for experiments which are only sensitive to low multipoles.

Although the previous theorem is mainly useful in the context of perturbative deformations of GR black holes, we should emphasize that all our results hold non-perturbatively in the background metric, subject only to the analyticity assumption \eqref{eq:general metric ansatz}. This fact, we stress, makes our method qualitatively different from other analytical approaches to calculate Love numbers in spacetimes beyond GR. As an illustration, we consider here two well-known examples of regular black holes~\cite{Bardeen:1968a,Hayward:2005gi}:\footnote{See~\cite{Garcia-Saenz:2025urd,Wang:2025oek,Bhattacharyya:2026itm} for recent work on Love numbers of regular black holes.}
\beq\bal \label{eq:hayward bardeen}
\,&\mbox{Hayward:}\qquad f=g=1-\frac{x}{1+\eta x^3} \,,\\
\,&\mbox{Bardeen:}\qquad f=g=1-\frac{x}{(1+\chi x^2)^{3/2}} \,,\\
\eal\eeq
where $0<\eta,\chi<\tfrac{4}{27}$. Table \ref{tab:regular bh} displays the \textit{exact} results of evaluating the static and lowest-order dynamical logarithmic Love numbers $\alpha_0$ and $\alpha_1$ for the lowest multipoles. For the Hayward metric, the vanishing of $\alpha_0$ at $\ell=1$ is in agreement with the above theorem. Similarly, the fact that $\alpha_1$ at $\ell=0$ does not receive corrections beyond the Schwarzschild value is a general prediction.
\begin{table}[h]
	\begin{tabular}{lcccc}
		\toprule
		&            & $\ell=0$ & $\ell=1$ & $\ell=2$ \\ \hline
		\multirow{2}{*}{Hayward} & $\quad\alpha_0\quad$ & $0$ & $0$ & $\tfrac{4\eta}{5}$        \\ 
		& $\quad\alpha_1\quad$ & $1$ & $\tfrac{5+8\eta}{60}$ & $\tfrac{49+5926\eta-252\eta^2}{8820}$       \\ \hline
		\multirow{2}{*}{Bardeen} & $\quad\alpha_0\quad$ & $0$ & $-\tfrac{\chi}{2}$ & $\tfrac{-3\chi+15\chi^2}{10}$      \\ 
		& $\quad\alpha_1\quad$ & $\tfrac{2 - 3\chi}{2}$ & $\tfrac{50-663\chi+150\chi^2}{600}$ & $\tfrac{196-13152\chi+73992\chi^2+2205\chi^3}{35280}$     \\
		\bottomrule
	\end{tabular}
	\caption{Logarithmic Love numbers $\alpha_0$ and $\alpha_1$ for the Hayward and Bardeen black holes defined by the metric functions in \eqref{eq:hayward bardeen}.}
	\label{tab:regular bh}
\end{table}

Perhaps more interesting is to ask about scheme-invariant quantities such as $\delta^2$. As an example, we display here the $\ell=0$ result for the Bardeen metric:
\beq
(\delta^{(\ell=0)})^2=\left(\frac{49}{36} + 3\chi\right)\omega^4+ \left(\frac{9449}{3240} + \frac{209\chi}{20} + \frac{37\chi^2}{12}\right)\omega^6+\mathcal{O}(\omega^8) \,.
\eeq
This shows that $\delta^2$ can indeed become negative, as discussed in Sec.\ \ref{sec:monodromy}. However this does not occur for a regular black hole, which requires $0<\chi<\tfrac{4}{27}$. We have checked that the same conclusion applies for $\ell=0,1,2$ scalar perturbations for both the Hayward and Bardeen metrics.


\section{Discussion} \label{sec:conclusion}

In this paper we aimed at clarifying that the running of dynamical Love numbers is not given by an infinite collection of independent quantities but the exponential of a single $2\times2$ matrix, and that the three entries of that matrix --- the running of the Love number $A$, the anomalous dimension $C$ and the tail-like term $B$ --- are computable by a finite algebraic recursion on any static spherically symmetric background and for any type of perturbation. The theorem behind this statement is elementary: it is the monodromy structure of Fuchsian theory applied to a tower of sourced equations. 

It would be worthwhile to make the physical interpretation of the three logarithmic coefficients more precise, for instance by reproducing our universal result for $\beta_{\ell+1}$ in the world-line EFT, and more generally to match $A$, $B$ and $C$ to the EFT beta functions order by order. We have done so only at the level of the invariant $\delta^2$ for the monopole scalar perturbation on Schwarzschild, where the agreement with the $\overline{\rm MS}$ results of~\cite{Caron-Huot:2025tlq} gave a highly non-trivial check of both calculations, and where the individual coefficients differ exactly as predicted by the covariance of the monodromy generator. Extending this comparison to all multipoles and spins, and to the Kerr results obtained from scattering amplitudes~\cite{Saketh:2023bul,Ivanov:2024sds,Bautista:2023sdf}, would provide a valuable cross-check.

We think the lessons on the issue of scheme dependence deserve to be reiterated. The static logarithmic Love number $\alpha_0$ is universal, while the leading dynamical term, $\alpha_1$, is universal only if $\alpha_0=0$ \textit{and} the static basis of solutions is fixed. The dipole electromagnetic perturbation of RN provides an instructive example where $\alpha_0=0$ and yet $\alpha_1$ is ambiguous, because the horizon-regular static solution happens to reach the resonant power and the static Love number $\mathcal{K}_0$ is itself scheme-dependent. The moral is that one should either spell out the scheme in which a coefficient is quoted, or else to focus on scheme-invariant quantities such $\delta^2$.

Several extensions and generalizations suggest themselves. It would be straightforward to study the Zerilli and the even-parity Moncrief equations with our method. The obvious more interesting extension is the case of spinning black holes. The radial Teukolsky equation has the same singularity structure at infinity and therefore the near-zone expansion in the frequency should proceed in the same way. The new ingredients we foresee are the frequency dependence of the angular eigenvalue and the loss of the $\omega\to-\omega$ symmetry, so that the tower is expected to be organized in powers of $\omega$ rather than $\omega^2$, but the inductive proof of the theorem does not rely on this property and we expect the monodromy generator to retain its general structure. Coupled systems of perturbations, as generically occur in the context of black holes beyond GR~\cite{Barbosa:2026qcv,Becar:2026zqd}, can also be studied with our techniques. With $N$ coupled equations, the monodromy generator is a $2N\times2N$ matrix, and one expects the renormalization group flow to be determined by a matrix equation, but otherwise with the same general structure as outlined here.

Finally, the vanishing of the leading dynamical running for extremal RN black holes, proved here for all multipoles in the axial sector, enlarges the list of special properties of Love numbers of extremal black holes~\cite{Noumi:2026shc,Xia:2025zfp,Kehagias:2024yzn,Pang:2025myy,Pang:2026vah,Gounis:2025tmt}. Its mechanism appears to be transparent enough through our residue formula and the fact that the static tidal field vanishes on the degenerate horizon, yet it is natural to ask whether this might be related to a symmetry, in the spirit of the ladder symmetries for static Love numbers (which is hinted at by our results for the anomalous dimension coefficients), or to some deeper property of extremal black holes. It would be worthwhile to check the polar sector as well as higher-order scheme-invariant dynamical coefficients. We plan to address some of these questions in a forthcoming work.

\subsubsection*{Acknowledgments}

The author would like to thank Giorgio di Russo, Hongbo Lin, Anna Tokareva and Jun Wang for useful conversations. The author acknowledges support from a Provincial Grant (Grant No.\ 2023QN10X389).

\appendix

\section{Leading tail term derivation} \label{app:beta universal}

We wish to compute $\beta_{\ell+1}\equiv D_{\ell+1,1,1}$ from the master recurrence formula for $\ell\geq 1$. The decaying branch exhibits a resonance at $n=1$ at level $k=\ell+1$, so that $\beta_{\ell+1}$ is determined in terms of $D_{\ell+1,1-m,j}$ with $m\geq1$ and in terms of $D_{\ell,1-m,j}$ with $m\geq0$. It follows that $\beta_{\ell+1}$ can only a priori depend on $p_1$, $q_1$, $w_0$ and $w_1$. We will see that, in fact, it suffices to know $w_0=1$ and $w_1=2$.

We must first solve the recurrence for the `sub-hierarchy' $n\leq1$ and $1\leq k\leq\ell$. Notice that this set of $(k,n)$ does not hit any resonances and has no logarithms to begin with at level 0. Therefore no logarithms can appear in this sub-hierarchy and only $D_{k,n,0}$ is non-zero.
For $j=0$, $n=0$ the recurrence reads $I(\ell-2k)D_{k,0,0}+D_{k-1,0,0}=0$. The solution is
\beq
D_{k,0,0}=\frac{(2\ell-2k-1)!!}{2^kk!(2\ell-1)!!} \,,
\eeq
valid for $0\leq k\leq \ell$ and with the normalization $D_{0,0,0}\equiv 1$. For $j=0$, $n=1$ the recurrence reads $I(\ell-2k+1)D_{k,1,0}+[(\ell-2k)p_1+q_1]D_{k,0,0}+D_{k-1,1,0}+2D_{k-1,0,0}=0$ for $0\leq k\leq\ell$. This can be rewritten as
\beq
\eta_k=\frac{-[(\ell-2k)p_1+q_1]+I(\ell-2k)(\eta_{k-1}+2)}{I(\ell-2k+1)} \quad (1\leq k\leq\ell) \,,\qquad \eta_0=-\frac{\ell p_1+q_1}{2(\ell+1)} \,,
\eeq
where $\eta_k:= D_{k,1,0}/D_{k,0,0}$. Since this recurrence is linear in $p_1$ and $q_1$, the solution must have the form $\eta_k=\eta^{(p)}_kp_1+\eta^{(q)}_kq_1+2\tilde{\eta}_k$, where each term obeys an independent recursion. We obtain\footnote{The solution for $\tilde{\eta}_k$ can be expressed as a hypergeometric series, however here we shall only need the specific value $\tilde{\eta}_\ell$.}
\beq
\eta^{(p)}_k=-\frac{\ell^2+2k}{2\ell(\ell+1)} \,,\qquad \eta^{(q)}_k=\frac{2k-\ell}{2\ell(\ell+1)} \,,\qquad \tilde{\eta}_\ell=\frac{4^{\ell}(\ell!)^2}{(2\ell)!}-1 \,.
\eeq

Lastly we glean at the resonant level $k=\ell+1$, $n=1$, $j=0$: $-(2\ell+1)\beta_{\ell+1}+[(-\ell-2)p_1+q_1]D_{\ell+1,0,0}+D_{\ell,1,0}+2D_{\ell,0,0}=0$. Remarkably, we find that $p_1$ and $q_1$ cancel upon substituting the above results. What remains is
\beq
(2\ell+1)\beta_{\ell+1}=2D_{\ell,0,0}(\tilde\eta_{\ell}+1) \,,
\eeq
which simplifies to the result quoted in Sec.\ \ref{sec:black holes}.

\section{Leading dynamical Love running for the RN black hole} \label{app:alpha1 RN}

\textbf{Case $\ell\geq2$:} The proof of Eq.\ \eqref{eq:a1-RN} follows most directly from the residue formula \eqref{eq:residue2}. The solutions of the static Moncrief equations are indeed Laurent polynomials, which is the very reason why the static Love numbers are zero. The complication is that these equations are of the Heun type, making it difficult to establish closed-form results valid for arbitrary $\ell$ (although for any given $\ell$ running the master recurrence formula is straightforward).

We can however bypass this issue by means of a change of variable that turns the Moncrief equations into hypergeometric form~\cite{Rai:2024lho}. Thanks to this trick we will prove that
\beq\bal \label{eq:rn horizon values}
u_{\rm g}^{(0)}(r_{+})&=\frac{(\ell-2)!(\ell+1)!}{(2\ell)!}\,r_+(r_+-r_-)^{\ell-1}\left(\mathcal{Q}_{\pm}+\ell-1\right) \,,\\
u_{\rm g}^{(0)}(r_{-})&=(-1)^{\ell+1}\frac{(\ell-2)!(\ell+1)!}{(2\ell)!}\,r_-(r_+-r_-)^{\ell-1}\left(\mathcal{Q}_{\pm}+\ell-1\right) \,,
\eal\eeq
the static growing solutions evaluated at the RN horizons $r_{\pm}$ (notice the sign in $r_{\pm}$ is unrelated to the sign in $\mathcal{Q}_{\pm}$, the latter labeling the two Moncrief equations) and normalized such that $u_{\rm g}^{(0)}\sim x^{-\ell-1}$ at $x=0$.

The master equation of~\cite{Rai:2024lho} is
\begin{equation}
	z(1-z)F''+(1+2z)F'+(\ell+2)(\ell-1)F=0 \,,
	\label{eq:RS-hyper}
\end{equation}
in terms of the variable $z:=(r-r_-)/S$, $S:=r_+-r_-$. Its solution is
\begin{equation}
F=\sum_{n=0}^{\ell+2}F_n z^n \,,\qquad F_n=(-1)^n\binom{\ell+2}{n}\binom{\ell-2+n}{\ell-2} \,,
\label{eq:RS-F}
\end{equation}
a polynomial of degree $\ell+2$. In particular, the leading coefficient at spatial infinity is $F_{\ell+2}=(-1)^\ell\tfrac{(2\ell)!}{(\ell-2)!(\ell+2)!}$, and it is straightforwardly checked that
\begin{equation}
	F(1)=F'(1)=F''(1)=F'''(1)=0 \,,\qquad F''''(1)=(-1)^\ell\frac{(\ell+2)!}{(\ell-2)!} \,,
	\label{eq:RS-zero}
\end{equation}
at the outer horizon $z=1$.

The relation between the Moncrief variable $u$ and $F$ is
\beq
u_{\rm g}^{(0)}\propto \frac{1}{r}\mathcal{P}\left(z\frac{\d}{\d z}\right)F \,,\qquad \mathcal{P}(X):=X^4+aX^3+bX^2+cX+d \,,
\eeq
for some known coefficients $a,b,c,d$, given explicitly in~\cite{Rai:2024lho} (notice that these carry an implicit sign label $\pm$ for each of the Moncrief variables). We fix the proportionality constant such that $u_{\rm g}^{(0)}\sim r^{\ell+1}$ at $r\to\infty$,
\beq
u_{\rm g}^{(0)}=\frac{S^{\ell+2}}{rF_{\ell+2}\mathcal{P}(\ell+2)}\mathcal{P}\left(z\frac{\d}{\d z}\right)F \,.
\eeq
Thanks to \eqref{eq:RS-zero}, at the outer horizon we simply have $\mathcal{P}\left(z\frac{\d}{\d z}\right)F\big|_{r=r_+}=F''''(1)$, given above. It is tedious but straightforward to evaluate the constant $\mathcal{P}(\ell+2)$. Doing this and simplifying eventually yields $u_{\rm g}^{(0)}(r_{+})$ as written in \eqref{eq:rn horizon values}.

The result of evaluating at the inner horizon is immediately inferred from the observation that \textit{each} Moncrief equation is manifestly symmetric under the exchange $r_+\leftrightarrow r_-$ (which sends $S\to-S$; notice that $\mathcal{P}(\ell+2)$ carries an extra factor $S$, hence the coefficient $(-1)^{\ell+1}$ in \eqref{eq:rn horizon values}).

\textbf{Case $\ell=1$:} The dipole $\ell=1$ must be treated separately. It is also quite instructive. The `$+$' Moncrief equation is then unphysical, corresponding to a dipole gravitational mode. The `$-$' equation has the exact horizon-regular solution (returning here to our standard variable $x$)
\beq
u^{(0)}=\frac{1}{x^2}+3f_1+4f_1^2x \,.
\eeq
This shows that, \textit{in the pure-power scheme}, the static Love number is non-zero:
\beq
\mathcal{K}_0^{(\ell=1)}=4f_1^2 \,.
\eeq
Notice that the running is still zero, $\alpha_0=0$, and this \textit{is} a scheme-invariant statement. At leading-order dynamical level we find
\beq
\alpha_1=\frac{66f_1^2+50f_1+5}{15} \,,
\eeq
which however is not scheme-invariant. Under a static shift $u_{\rm g}^{(0)}\mapsto u_{\rm g}^{(0)}+cu_{\rm d}^{(0)}$, as discussed in Sec.\ \ref{sec:monodromy}, we have
\beq \label{eq:scheme change RN}
\mathcal{K}_0\mapsto \mathcal{K}_0-c \,,\qquad \alpha_1\mapsto \alpha_1+2c\gamma_1 \,,\qquad \gamma_1\mapsto \gamma_1 \,,
\eeq
so that the relevant scheme-invariant quantity is
\beq
\tilde{\alpha}_1:=\alpha_1+2\gamma_1\mathcal{K}_0 =\frac{(1+2f_1)(1+4f_1)^2}{3} \,,
\eeq
which is the result we quoted in the main text.

\textbf{Scheme invariance for $\ell\geq2$:} Having understood the scheme dependence issue for $\ell=1$, we can return to the generic $\ell\geq2$ case and ask the same question. Notice that \eqref{eq:scheme change RN} holds generally for all $\ell\geq1$ (cf.\ Sec.\ \ref{sec:monodromy}). However, the static Love number $\mathcal{K}_0$ does vanish in the pure-power scheme for $\ell\geq2$ in RN, so that $\alpha_1=\tilde{\alpha}_1$ in this scheme.

It is instructive to see why the case $\ell=1$ is special in this regard. Suppose we know that the physical, horizon-regular static solution is a Laurent polynomial,
\beq
u^{(0)}=x^{-\ell-1}\left(1+u_1x+\cdots +c_Nx^N\right) \,.
\eeq
Clearly, barring an accidental cancellation, $u^{(0)}$ will contain the power $x^{\ell}$ whenever $N\geq 2\ell+1$, and thus a non-zero Love number (in the pure-power scheme). $N$ of course is determined by the equation, and for the Moncrief equations we know it is $N=\ell+2$ from \eqref{eq:RS-F}. Therefore $N\geq 2\ell+1$ happens precisely only for $\ell=1$.

\section{Proof of the vanishing Love running theorem} \label{app:zero love theorem}

Let $f=g=1-x\left(1+ax^N\right)$. Let $F$ be any power series in $x$ in the problem, where we assume we have factored out powers and logarithms such that the series starts at $\mathcal{O}(x^{0})$, i.e.\ it is (formally) a Taylor series. Consider any Taylor coefficient $Y$ in $F$ and expand it as a (formal) power series in $a$: $Y=\sum_{\sigma\geq0}Y^{(\sigma)}a^\sigma$. Let $\mathcal{S}_N$ be the set of coefficients such that $Y\in\mathcal{S}_N$ means that $Y^{(\sigma)}=0$ whenever the power of $x$ it multiplies is less than $\sigma(N+1)$. We will also say that $F\in\mathcal{S}_N$ if all its Taylor coefficients are in $\mathcal{S}_N$. This means that the Taylor series $F[a^\sigma]$ must start at $\mathcal{O}(x^{\sigma(N+1)})$, at least. (Here $F[a^\sigma]$ denotes the coefficient of $a^\sigma$ in $F$.)

Our main claim is: $D_{k,n,j},G_{k,n,j}\in\mathcal{S}_N$ for all $k,n,j$. Equivalently, since for these coefficients $n$ labels the power of $x$ in the above definition, then $D_{k,n,j}^{(\sigma)},G_{k,n,j}^{(\sigma)}=0$ for all $n<\sigma(N+1)$. The theorem of Sec.\ \ref{sec:black holes} then immediately follows, because $\alpha_k\equiv -G_{k,2\ell+1+2k,1}$, $\gamma_k\equiv G_{k,2k,1}$ and $\beta_k\equiv D_{k,2k-2\ell-1,1}$.

Next we prove the above claim. Consider first the functions $p$, $q$ and $x^2\mathcal{R}$ which define the ODE. Since they are analytic by assumption, they are of the form in the above definition, and it is easy to see that they are in $\mathcal{S}_N$, as can be checked term by term.\footnote{In fact it suffices to check explicitly $1/f$ (the easiest case). Then the statement follows for $1/f^2$ and $xf'/f$ because $\mathcal{S}_N$ is closed under addition, multiplication and the action of $x\frac{\d}{\d x}$.} Let
\beq \label{eq:app Tdef}
T_{k,n,j}:= \sum_{m=1}^{n}\Big\{\big[(s-m)p_m+q_m\big]E_{k,n-m,j}+(j+1)p_mE_{k,n-m,j+1}\Big\}
+\sum_{m=0}^{n}w_mE_{k-1,n-m,j} \,,
\eeq
so that the master recurrence formula is $I(s)E_{k,n,j}+(j+1)I'(s)E_{k,n,j+1}+(j+1)(j+2)E_{k,n,j+2}=-T_{k,n,j}$, and here $E$ stands for either $G$ or $D$.

We now use induction to argue that $T_{k,n,j}\in \mathcal{S}_N$. This is true for $k=0$ by the theorem proved in~\cite{Garcia-Saenz:2022wsl}. Assume next it holds up to level $k-1$. The last term in \eqref{eq:app Tdef} is then clearly in $\mathcal{S}_N$. The master recursion for $n=0$ then says that $I(s)E_{k,0,j}+(j+1)I'(s)E_{k,0,j+1}+(j+1)(j+2)E_{k,0,j+2}$ is in $\mathcal{S}_N$, and therefore $E_{k,0,j}\in \mathcal{S}_N$ because multiplication by logarithms do not affect this property (notice that $I(s)$ is independent of $a$). Next we use induction over $n$: we have just stated the base case $n=0$, and we assume the statement holds up to $n-1$. This proves that the first sum in \eqref{eq:app Tdef} is in $\mathcal{S}_N$ and therefore $T_{k,n,j}\in \mathcal{S}_N$. Finally, returning to the master recursion, we conclude that $E_{k,n,j}\in \mathcal{S}_N$.

We remark that there is a hidden assumption we have made: to conclude that the master formula maintains the property of being in $\mathcal{S}_N$ at each recursion, we must assume that the initial conditions are independent of $a$. This is true in our unit pure-power scheme: we start with $E_{0,0,0}\equiv 1$, and at each resonance that leaves some $E_{k,n,0}$ undetermined, we set it to zero.


\bibliographystyle{JHEP}
\bibliography{LoveMonodromy_refs}

\end{document}